\documentclass[12pt,aps,superscriptaddress,preprintnumbers,nofootinbib]{revtex4-1} 
\pdfoutput=1
\usepackage{amsmath,amssymb,graphicx,natbib,bm}
\usepackage{enumerate}
\usepackage{hyperref}
\usepackage{color}
\usepackage{url}
\usepackage{multirow}

\usepackage{empheq}
\usepackage{slashed}

\newcommand{\beq}{\begin {equation}}  
\newcommand{\eeq}{\end   {equation}} 
\newcommand{\bea}{\begin {eqnarray}} 
\newcommand{\eea}{\end   {eqnarray}}  
\newcommand{\baa}{\begin {array}   } 
\newcommand{\eaa}{\end   {array}   }     
\newcommand{\bit}{\begin {itemize} }
\newcommand{\eit}{\end   {itemize} }
\newcommand{\be }{\begin {equation}} 
\newcommand{\ee }{\end   {equation}}
\newcommand{\nn }{\nonumber        }

\newcommand{\GeV}{\ensuremath{\mathrm{GeV}}}

\begin{document}


\preprint{ACFI-T17-23}

\title{Implications of hidden gauged $U(1)$ model for $B$ anomalies}

\author{Kaori Fuyuto}
\email{kfuyuto@umass.edu}
\author{Hao-Lin Li}
\email{haolinli@physics.umass.edu}
\author{Jiang-Hao Yu}
\email{jhyu@physics.umass.edu}
\affiliation{Amherst Center for Fundamental Interactions, Department of Physics, University of Massachusetts-Amherst, Amherst, MA 01003, U.S.A.}


\begin{abstract}

We propose a hidden gauged $U(1)_H$ $Z'$ model to explain deviations from the Standard Model (SM) values in lepton flavor universality known as $R_K$ and $R_D$ anomalies.  
The $Z'$ only interacts with the SM fermions via their mixing with vector-like doublet fermions after the $U(1)_H$ symmetry breaking, which leads to $b \to s \mu\mu$ transition through the $Z^{\prime}$ at tree level.
Moreover, introducing an additional mediator, inert-Higgs doublet, yields $b\to c \tau \nu$ process via charged scalar contribution at tree level. Using flavio package, we scrutinize adequate sizes of the relevant Wilson coefficients to these two processes by taking various flavor observables into account. It is found that significant mixing between the vector-like and the second generation leptons is needed for the $R_K$ anomaly.
A possible explanation of the $R_D$ anomaly can also be simultaneously  addressed in a motivated situation, where a single scalar operator plays a dominant role, by the successful model parameters for the $R_K$ anomaly.

\end{abstract}

\maketitle


\section{Introduction}
\label{sec:intro}


While there is null result from direct searches for new physics at the Large Hadron Collider, recent LHCb measurements of lepton flavor universality in $B$ physics have shown deviations from the standard model (SM) predictions. 
The measurement of $R_{K^{*0}} = {\mathcal B}(B^0 \to K^{*0} \mu^+\mu^-)/{\mathcal B}(B^0 \to K^{*0} e^+e^-)$ in two different kinematic regions by LHCb collaboration gives~\cite{Aaij:2017vbb}:
\begin{eqnarray}
R_{K^{*0}}  = \left\{\begin{array}{ll}
  0.66^{+0.11}_{-0.07} \pm 0.03 & (2m_\mu)^2 < q^2 < 1.1~ \GeV^2  \\[2mm]
  0.69^{+0.11}_{-0.07} \pm 0.05& 1.1 ~ \GeV^2 < q^2 < 6 ~  \GeV^2 .
\end{array}
\right.
\end{eqnarray}
These values are compatible with the SM predictions \cite{Bordone:2016gaq,Descotes-Genon:2015uva,Capdevila:2016ivx,Capdevila:2017ert,Serra:2016ivr,Dvan,Straub:2015ica,Altmannshofer:2017fio,flavio,Jager:2014rwa} within $2.1-2.3\sigma$ and $2.4-2.5\sigma$, respectively.
The ratio of $R_K={\mathcal B}(B^+ \to K^{+} \mu^+\mu^-)/{\mathcal B}(B^+ \to K^{+} e^+e^-)$ is also given by LHCb collaboration \cite{Aaij:2014ora} which shows 2.6$\sigma$ deviation from the SM predictions \cite{Bordone:2016gaq}: 
\begin{eqnarray}
R_{K}  = \begin{array}{ll}
  0.745^{+0.090}_{-0.074} \pm 0.036& 1 ~ \GeV^2 < q^2 < 6 ~  \GeV^2 .
\end{array}
\end{eqnarray}
Furthermore, charged current decays of $\bar{B}\to D^{(*)}l^-\bar{\nu}_l$, which have been measured by the BaBar \cite{Lees:2012xj,Lees:2013uzd}, Belle \cite{Huschle:2015rga, Sato:2016svk, Hirose:2016wfn} and LHCb \cite{Aaij:2015yra}, also indicate discrepancies in the ratios of $R_{D^{(*)}}={\mathcal B}(\bar{B}\to D^{(*)}\tau^-\bar{\nu}_{\tau})/{\mathcal B}(\bar{B}\to D^{(*)}l^-\bar{\nu}_{l})~(l=e,\mu)$. The SM predictions are estimated in \cite{Fajfer:2012vx,Bigi:2017jbd,Jaiswal:2017rve} for $R(D^*)$ and \cite{Lattice:2015rga,Na:2015kha,Bigi:2016mdz} for $R(D)$, and the current experimental average values are roughly $4\sigma$ deviations from the SM values.
In recent years, these anomalies have been paid more attentions, and possibilities for the explanation in various extensions of the SM are discussed: $Z'$ models \cite{Altmannshofer:2016jzy,Ko:2017lzd,Datta:2017pfz,Datta:2017ezo, Bhattacharya:2016mcc, Cline:2017ihf,Faisel:2017glo}, leptoquark \cite{Bhattacharya:2016mcc, Hiller:2014yaa, Duraisamy:2016gsd, Chen:2017hir, Aloni:2017ixa, Calibbi:2017qbu} for the $R_K$ anomaly, and the charged Higgs \cite{Bailey:2012jg, Celis:2012dk, Tanaka:2012nw, Crivellin:2012ye, Crivellin:2015hha, Cline:2015lqp, Alonso:2016oyd, Iguro:2017ysu}, $W^{\prime}$ models \cite{He:2012zp,Boucenna:2016qad} and leptoquark \cite{Chen:2017hir, Calibbi:2017qbu, Fajfer:2012jt, Sakaki:2013bfa, Dorsner:2013tla, Li:2016vvp, Crivellin:2017zlb} for the $R_{D}$ anomaly.

We propose a hidden gauged $U(1)_H$ model to explain the $R_{K}$ anomaly, in which a $Z'$ gauge boson  interacts with the SM particles only through mediator particles. The mediators are vector-like doublet fermions, whose masses are assumed to be around $O(1)~$TeV. The $Z^{\prime}$ couplings to the SM particles appear with symmetry breaking of the hidden gauged $U(1)_H$, which results in  $b \to s \mu\mu$ transition through the tree-level $Z^{\prime}$ exchange. Furthermore, introducing an another mediator, we can have a relevant $b\to c$ transition to the $R_D$ anomaly. The role of the additional mediator is played by an inert-Higgs doublet, and the charged Higgs component of the inert Higgs can induce the $b\to c$ transition at tree level. Since some essential couplings for the $b\to s \mu\mu$ are also involved in $b\to c \tau^- \bar{\nu}$, the former process can affect a possibility of explaining the $R_D$ anomaly. 

Typically, the $b\to s$ transition is tightly constrained by various observables such as $B_s-\bar{B}_s$, $B_s \to \mu\mu$, etc. While in the lepton sector, the $Z^{\prime}$ coupling to the muon contributes to neutrino trident production as discussed in \cite{Altmannshofer:2014cfa, Altmannshofer:2014pba}. 
At the same time, we carefully examine that there is no significant FCNC $Z$ couplings, although the mixing between the mediator and the SM fermions could be significantly large. In addition to these constraints, we also address direct searches of the $Z'$ and the charged Higgs at collider experiments.  
To avoid flavor constraints from the neutral Higgses, we take degenerate masses on the inert Higgs spectra. 
We utilize flavio package \cite{flavio} to perform a comprehensive analysis on various flavor observables, and obtain the global fit to the Wilson coefficients on the $R_K$ and $R_D$ anomalies. 
Applying these results to our hidden gauged model, we scrutinize the favored parameter space on explaining these anomalies.

The structure of this paper is as follows. We introduce the hidden gauged extension of the SM in Sec. II. In Sec. III, the mixing between the vector-like and SM fermions and the $Z^{\prime}$ couplings are presented.  
Using the flavio, we perform a global fit on the relevant Wilson coefficients to the $R_K$ and $R_D$ anomalies in Sec. IV. 
Subsequently, various experimental searches and constraints are discussed in section VI. In Sec. V, we apply the obtained results in previous sections to the model parameters. Finally, Sec. VII is devoted to conclusions.


\section{The Model}
\label{sec:model}

\begin{figure}[t]
\begin{center}
\includegraphics[width=10cm]{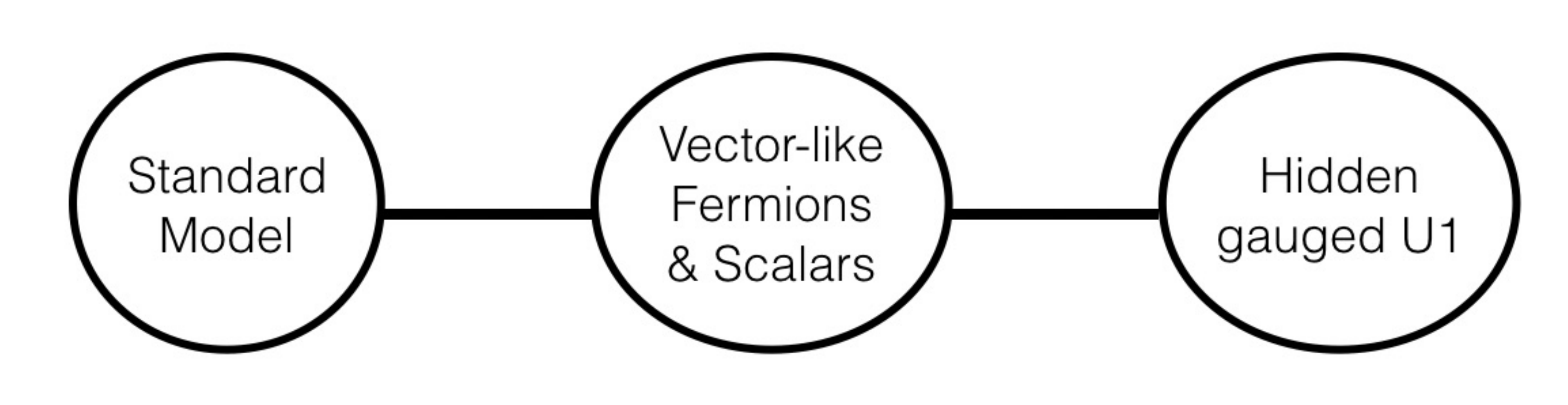} 
\end{center}
\caption{The schematic description of the hidden gauged $U(1)_H$ model.}
\label{fig:model}
\end{figure}

We consider a hidden sector extension of the SM under the hidden gauged $U(1)_H$ symmetry~\cite{Yu:2016lof}. Although all the SM particles are not charged under the $U(1)_H$ symmetry, the $U(1)_H$ gauge boson can couple to them through mediators, such as new vector-like fermions or scalars. The mediators have both the $U(1)_H$ and the SM gauge charges. The schematic framework of this model is described in Fig.~\ref{fig:model}.

For the $R_K$ anomaly, essential mediators are the vector-like fermions. These fermions 
could be $SU(2)_L$ singlet, doublet, or multiplets, with one or more generations. 
Here, we focus on one minimal fermion assignments: all the vector-like fermions are 
$SU(2)_L$ doublets, with only one generation~\footnote{
Another minimal choice is to introduce only vector-like singlet fermions with one generation. 
Since our purpose is to illustrate how the $U(1)_H$ works for the $R_K$ anomaly, the 
fermion assignment is not so important. We expect that the analysis is quite similar to different fermion assignments.}. 
To be free of gauge anomaly, the new vector-like fermions need to possess appropriate quantum numbers. 
We assign the SM quantum numbers on the vector-like fermions such that they mix with the SM fermions at tree level.

Moreover, if we introduce additional mediators such as a charged scalar, 
the $R_D$ anomaly could also be explained as a bonus. 
In this work, we add a second scalar doublet $H^{\prime}$ as an inert type~\cite{Ma:2006km, Yu:2016lof} to keep our setup simple.

{\renewcommand{\arraystretch}{1.5} 
\begin{table}[ht]
\begin{center}
\begin{tabular}{c| c c c | c }
\hline \hline 
      &  $SU(3)_C$  & $SU(2)_L$ & $U(1)_Y$ & $U(1)_{H}$ \\ 
\hline
$ Q_{L,R}=(\tilde{U}_{L,R}~\tilde{D}_{L,R})^T$          & {\bf 3 }    &  {\bf 2}         &$+\frac{1}{6}$        & $+1$     \\
$ L_{L,R}=(\tilde{N}_{L,R},~\tilde{L}_{L,R})^T$          & {\bf 1 }    &  {\bf 2}         &$-\frac{1}{2}$        & $+1$    \\
\hline
$\Phi$         & {\bf 1 }    &  {\bf 1}         &$ 0$                  & $+1$           \\
\hline 
$H'$ (optional)          & {\bf 1 }    &  {\bf 2}         &$ \frac{1}{2}$        & $+1$   \\
\hline 
$\chi$ (optional)          & {\bf 1 }    &  {\bf 1}         &$0$        & $-1$   \\
\hline \hline 
\end{tabular}
\end{center}
\caption{
The new particle contents and their quantum numbers in the gauge group $SU(2)_L \times U(1)_Y \times U(1)_{H}$. 
Here $Q_{L,R}$ and $L_{L,R}$ denote the vector-like representation under the gauge group.
All the SM particles are not charged under $U(1)_{H}$.  The $H'$ doublet is optional: 
without $H'$ added, only $R_K$ anomaly is addressed; with $H'$, we could also explain the $R_D$ anomaly.
The vector-like fermion $\chi$ is also optional, which is $Z_2$-odd under $U(1)_{H} \to Z_2$ symmetry breaking,
and thus it is the dark matter candidate.
}
\label{table1}
\end{table}
 
The new particle contents and their quantum numbers are shown in Tab.~\ref{table1}. 
Under this charge assignments, the model is free of gauge anomaly.
The $U(1)_H$ symmetry spontaneously breaks when a singlet scalar $\Phi$, which is only charged under the $U(1)_H$, obtains a nonzero vacuum expectation value (VEV) $v_{\Phi}$. They are cast into the form
\bea
\Phi = \frac{s + i a}{\sqrt2}, \hspace{1cm} \langle \Phi \rangle = \frac{v_{\Phi}}{\sqrt{2}},
\eea
and the Lagrangian is written as
\bea
	{\mathcal L}_{\rm kin} =  D_\mu \Phi^* D_\mu\Phi - V(H, \Phi, H'),
\eea 
with the covariant derivative $D_\mu \Phi = (\partial_\mu + i g'  Z'_\mu)\Phi$.
Here, $g'$ is the gauge coupling of the $U(1)_H$, and the gauge boson mass is given by $m_{Z'}=g'v_{\Phi}$. In the above potential, $H$ describes the SM $SU(2)$ doublet, and the two Higgs doublets are parametrized by
\bea
	H = \left(\begin{array}{c} G^+ \\ \frac{v + h + i G^0}{\sqrt2}\end{array}\right), \hspace{2cm}
	H' = \left(\begin{array}{c} H^{\prime +} \\ \frac{H' + i A'}{\sqrt2}\end{array}\right).
\eea
The potential $V(H, \Phi, H')$ is not relevant to our flavor study, and for simplicity, it is taken as
\bea
V(H, \Phi, H') &=& -\mu_{\Phi}^2 |\Phi|^2 + \lambda_{\Phi} |\Phi|^4 + \mu_{H^{\prime}}^2 |H^{\prime}|^2 + \lambda_{H^{\prime}} |H^{\prime}|^4  \nn 
	+  \lambda_{H\Phi} |H|^2|\Phi|^2  + \lambda_{H^{\prime}\Phi} |H'|^2 |\Phi|^2,
\eea
which respects a $Z_2(H') \times Z_2(\Phi)$ symmetry.  The $\mu_{H^{\prime}}^2$ parameter is positive, and thus $H^{\prime}$ does not get the VEV. 

The vector-like fermions are defined by
\bea
	Q_{L,R} = \left(\begin{array}{c} \tilde{U}_{L,R} \\ \tilde{D}_{L,R} \end{array}\right), \hspace{2cm}
	L_{L,R} = \left(\begin{array}{c} \tilde{N}_{L,R} \\ \tilde{L}_{L,R}\end{array}\right),
\eea
and their Dirac masses are  
\bea
	-{\mathcal L}_{\rm mass} = m_Q \bar{Q}_L Q_R  + m_L \bar{L}_L L_R +h.c..
\eea
All the new fermions are charged under $U(1)_H$, as shown in Table~\ref{table1}.\footnote{These new fermions contribute to the kinetic mixing between $U(1)_Y$ and $U(1)_H$ at the one loop order. The mixing parameter is estimated to be roughly $\frac{eg'}{(4\pi)^2} \ln \frac{m_Q}{m_L}$. Here for simplicity, we take small or zero mass splitting between the vector-like quarks and leptons, and thus the kinetic mixing is very tiny and negligible.}
The right-handed vector-like fermions can couple to the left-handed SM fermions through the  Yukawa interactions of $\Phi$
\bea
	-{\mathcal L}_{\rm Yukawa} &=&  \Phi^{\dagger}\sum_i\left( \overline{q}_{Li} Y_{q_iQ_R} Q_R  +  \overline{\ell}_{Li} Y_{l_iL_R} L_R\right) + {\rm h.c.},\nonumber\\
	&\supset &\frac{v_{\Phi}}{\sqrt{2}}\left( Y_{d_i\tilde D_R}\bar{d}_{Li}{\tilde D}_R
	+ Y_{u_i\tilde U_R}\bar{u}_{Li}{\tilde U}_R
	+Y_{l_i \tilde L_R}\bar{e}_{Li}\tilde{L}_R 
	+Y_{\nu_i \tilde N_R}\bar{\nu}_{Li}\tilde N_R+ {\rm h.c}.\right), 
\label{Yukawa_phi}
\eea
where the $SU(2)$ symmetry imposes a relation\footnote{If the vector-like fermions are the $SU(2)$ singlet, in general, they
introduce more free Yukawa couplings than the doublet case here. }
\bea
Y_{d_i\tilde D_R}=V^{\dagger}_{\rm CKM}Y_{u_i\tilde{U}_R}, \hspace{1cm}
Y_{l_i\tilde L_R}=V^{\dagger}_{\rm PMNS}Y_{\nu_i\tilde{N}_R} \label{Yukawa_relation}
\eea
with a generation index $i = 1,2,3$, Cabbio-Kobayashi-Maskawa (CKM) matrix $V_{\rm CKM}$ and Pontecorvo-Maki-Nakagawa-Sakata (PMNS) matrix $V_{\rm PMNS}$. Similar situations are studied in \cite{Altmannshofer:2014cfa, Fuyuto:2015gmk} for the quark sectors and \cite{Poh:2017tfo} for the lepton sectors. The interactions in Eq.~(\ref{Yukawa_phi}) describe the mixing between the vector-like and SM fermions, which yields $\bar{s}bZ^{\prime}$ and $\bar{\mu}\mu Z^{\prime}$ couplings after diagonalization of mass matrices. These $Z^{\prime}$ couplings are originated from three Yukawa components, $Y_{b\tilde D_R},~Y_{s\tilde D_R}$ and $Y_{\mu \tilde L_R}$.  

Apart from the right-handed vector-like fermions, the left-handed vector-like fermions have couplings to the right-handed SM fermions through the inert doublet
\bea
	-{\mathcal L}'_{\rm Yukawa} &=&  {\rm y}'_{Q_Lu_i} \overline{Q_L} \tilde{H}' u_R  +  {\rm y}'_{Q_Ld_i} \overline{Q_L} H' d_R 
	+ {\rm y}'_{L_Le_i} \overline{L_L} H' e_R  +{\rm  h.c.},
\label{Yukawa_H}
\eea
with $\tilde H^{\prime}=i\sigma_2H^{\prime *}$.
Unlike the Yukawa couplings in Eq.~(\ref{Yukawa_phi}), the above interactions do not contribute to the mass matrices of the fermions. $b\to c \tau \nu$ transition is induced by the charged scalar in the inert doublet, in which the vector-like fermions play a role in yielding $\bar{c}bH^{+}$ and  $\bar{\nu}\tau H^{+}$ couplings.

Since we introduce a hidden gauge sector, it is likely to have hidden particles, which are not charged under the SM gauge group at all. 
If so, the lightest un-colored one would be the dark matter candidate. 
Here we assume it is the WIMP dark matter. 
When we study the phenomenology of the hidden sector particles, typically only the lightest dark matter candidate plays an important role. 
For simplicity, we introduce a vector-like fermion $\chi$, which only carries the $U(1)_H$ charge
\bea
	{\mathcal L} = \bar{\chi} i\gamma^\mu D_\mu \chi, \quad D_\mu = \partial_\mu - i g' Z'_\mu.
\eea
To make it a dark matter candidate, we assign a $Z_2$ charge for the $\chi$. 
Under $U(1)_H$ breaking down to the $Z_2$ symmetry, the $\chi$ is $Z_2$-odd while all other mediators and the SM particles are $Z_2$ even.


\section{Mass Matrices and Couplings}
\label{sec:modelconstraint}
The interactions in Eq. (\ref{Yukawa_phi}) imply off-diagonal parts in fermion mass matrices, which are described by
\begin{eqnarray}
\left(\begin{array}{cc}
\overline{f}_{i,L} &\overline{\tilde{F}}_L 
\end{array}\right)
\left(
\begin{array}{cc}
\frac{v}{\sqrt{2}}y_{i} &\frac{v_{\Phi}}{\sqrt{2}}Y_{f_i,\tilde{F}_R} \\
0 & M_{F}
\end{array}
\right)
\left(
\begin{array}{c}
f_{i,R}
 \\
 \tilde{F}_R 
\end{array}
\right)+h.c. \equiv 
\overline{f}_{La} {\mathcal M}^f f_{Ra} + h.c.,
\end{eqnarray}
with $(f_{i},\tilde{F})^T_{L,R}=(u_{i},\tilde{U})^T_{L,R}$, $(d_{i},\tilde{D})^T_{L,R}$, $(l_{i},\tilde{L})^T_{L,R}$, $i=1, \dots, 3$ and $a = 1, \dots, 4$. $y_i$ is the SM Higgs Yukawa coupling, and its VEV is $v\simeq 246~$GeV. The matrix ${\cal M}^f$ is diagonalized by unitary matrices $U^f_L$ and $U^f_R$
\bea
	U_{L}^{f\dagger}  {\mathcal M}^f U_{R}^{f} = \textrm{diag}(m_{f_1},m_{f_2},m_{f_3},m_{f_4}),
\eea
which gives relationships between the original gauge eigenstate $f_a$ and new mass eigenstate $f^{\prime}_a$ as $f^{\prime}_{R}=(U^{f}_R)^{\dagger}f_{R}$ and $f^{\prime}_L=(U^f_L)^{\dagger}f_L$.
This diagonalization can affect original gauge interactions such as the $W$ and $Z$ boson couplings. However, it should be noted that, since the left-handed vector-like fermions have the same SM charges as those of the SM fermions, the $Z$ boson couplings to the left-handed fermions do not change. We will see this situation later.

In the charged-lepton sector, the mixing between the second generation lepton $\mu$ and vector-like lepton $\tilde{L}$ is essential for the $b\to s \mu\mu$ transition. Assuming there is only one mixing, namely, $Y_{\mu \tilde L_R}\neq 0$, the matrices $U_{L}^{\ell}$ and $U_{R}^{\ell}$ can be described by\footnote{Here, for simplicity, complex phases are not written down. }
\bea
	U_{L}^{\ell\dagger}  = 
	\left(\begin{array}{cccc}
	1 & 0  & 0 & 0 \\
	0 & c_{\alpha_L}  & 0 & -s_{\alpha_L} \\
	0 & 0  & 1 & 0 \\
	0 & s_{\alpha_L} & 0 & c_{\alpha_L}
	\end{array}
	\right)
	,\hspace{1cm}  U_{R}^{\ell}  = 
	\left(\begin{array}{cccc}
	1 & 0  & 0 & 0 \\
	0 & c_{\alpha_R}  & 0 & s_{\alpha_R} \\
	0 & 0  & 1 & 0 \\
	0 & -s_{\alpha_R} & 0 & c_{\alpha_R}
	\end{array}
	\right)
\eea 
where $s_{\alpha}\equiv \sin\alpha$ and $c_{\alpha}\equiv \cos\alpha$. The mixing angle $\alpha_L$ and $\alpha_R$ are approximately given by 
\bea
	s_{\alpha_L} \simeq \frac{ v_{\Phi}Y_{\mu \tilde{L}_R}/ \sqrt{2}  }{\sqrt{v^2_{\Phi}Y^2_{\mu \tilde{L}_R}/2 + M_L^2}}, \hspace{1cm}
s_{\alpha_R} \simeq \frac{m_\mu}{\sqrt{v^2_{\Phi}Y^2_{\mu \tilde{L}_R}/2 + M_L^2}} s_{\alpha_L}.
\eea
Since we suppose that $v_{\Phi}\sim O(100)~$GeV, $Y_{\mu\tilde L}\sim O(0.1-1)$ and  $M_{L}\sim O(1)~$TeV, it turned out that the angle $\alpha_L \gg \alpha_R$.

For the down-quark sector, the Yukawa couplings $Y_{b\tilde D_R}$ and $Y_{s\tilde D_R}$ are necessary for the $R_K$ anomaly, which implies the nonzero values of $Y_{t\tilde U_R}$ and $Y_{c\tilde U_R}$ with the same order of magnitude. Although, apart from the charged-lepton sector, the diagonalization is somewhat complicated due to the two mixing parameters, we obtain similar situation
\bea
	|1-U^{u\dagger}_{L} |_{ij} \gg |1-U^u_R|_{ij}, \hspace{1cm} |1-U^{d\dagger}_{L}|_{ij} \gg |1-U^{d}_{R} |_{ij}.
\eea
Again, the deviations of the right-handed mixing matrices from identity are suppressed by order $\frac{m_{q}}{m_Q}$.

For the neutrino parts, its Lagrangian contains
\begin{eqnarray}
{\cal L}_{\rm neutrino} \supset -M_L\overline{\tilde{N}}_L \tilde{N}_R - \frac{v_{\Phi}}{\sqrt{2}}Y_{\nu_i,\tilde{N}_R}\overline{\nu}_{Li} \tilde{N}_R +h.c. \label{neutrino}
\end{eqnarray}
As explained in the previous section, the second term is related to that of the charged lepton due to the $SU(2)$. As long as we keep only nonzero $Y_{\mu \tilde L_R}$, the  coupling is given by $Y_{\nu_i\tilde N_R}=(V_{\rm PMNS})_{i2}Y_{\mu  \tilde L_R}$. 
For sake of simplicity, we assume that the $V_{\rm PMNS}$ is a unit matrix, which results in only nonzero $Y_{\nu_{\mu}\tilde N_R}$ with more direct relationship of $Y_{\nu_{\mu}\tilde N_R}=Y_{\mu\tilde L_R}$. As in the same way of the charged-lepton sector, if we rotate the neutrino fields with
\bea
	\left(\begin{array}{c}
	\nu^{\prime}_{Li} \\
	\nu^{\prime}_{L4}
	\end{array}
	\right) = 
	U^{\nu\dagger}_L 
	\left(
	\begin{array}{c}
	\nu_{Li} \\
	\tilde{N}_L 
	\end{array}
	\right), \hspace{1cm}
	U^{\nu\dagger}_L = 	\left(\begin{array}{cccc}
	1 & 0  & 0 & 0 \\
	0 & c_{\beta_L}   & 0 & s_{\beta_L} \\
	0 & 0  & 1 & 0 \\
	0 & -s_{\beta_L}  & 0 & c_{\beta_L} 
	\end{array}
	\right)
\eea
where
\begin{eqnarray}
s_{\beta_L} = \frac{v_{\Phi}Y_{\mu \tilde{L}_R}/ \sqrt{2}  }{\sqrt{M_L^2+v_{\Phi}^2Y^2_{\mu,\tilde{L}_R}/2}} \simeq s_{\alpha_L}, \label{neutrino_mixing}
\end{eqnarray}
the Lagrangian in Eq. (\ref{neutrino}) becomes
\begin{align}
{\cal L}_{\rm neutrino}=-\sqrt{M^2_L+v^2_{\Phi}Y^2_{\mu\tilde L_R}/2}~\bar{\nu}^{\prime}_{L4}\tilde{N}_R+h.c.
\end{align}
We find that $U^{\nu}_L \simeq U_{L}^{\ell}$ due to the approximate $SU(2)$ symmetry, which deviates once the lepton masses are taken into account.

After diagonalizing the mass matrices, the $Z$ boson couplings to the lepton sector are 
\bea
	{\mathcal L}_Z \supset \frac{e}{s_W c_W} Z_\mu &\bigg[\bar{e}^{\prime}_{L} \gamma^\mu U^{\ell \dagger}_L(T^3_{L} + s_W^2 ) U^{\ell}_L e^{\prime}_{L} 
	+ \bar{e}^{\prime}_{R} \gamma^\mu U^{\ell \dagger}_R (T^3_{R} + s_W^2 ) U^{\ell}_R e^{\prime}_{R}\nonumber\\ 
	&- \bar{\nu}^{\prime}_L\gamma^{\mu}U^{\nu\dagger}_LT^3_{L}U^{\nu}_L\nu_L^{\prime} +\frac{1}{2} \bar{\tilde N}_R\gamma^{\mu}\tilde N_R
	\bigg],
\eea
where $T^3_{L}$ and $T^3_{R}$ are defined as $T^3_{L} = -\frac{1}{2} {\rm{diag}}(1,1,1,1)$ and $T^3_{R} =  -\frac{1}{2}{\rm{diag}}(0, 0,0 ,1) $. 
It is seen that the left-handed $Z$ couplings are diagonal, while the flavor-violating $Z$  couplings to the right-handed charged leptons exist. However, the right-handed mixing angle is tiny, which results in the small flavor-violating couplings. 
Similarly, in the quark sectors, while the $Z$ couplings to the left-handed quarks do not receive any corrections from the field definitions, the right-handed couplings have the flavor-violating parts but they are suppressed.

The $W$ couplings to the leptons are given by
\bea
	{\mathcal L}_W \supset \frac{e}{\sqrt2 s_W  } W^-_{\mu} \left[\bar{e}^{\prime}_{L} \gamma^\mu U^{\ell \dagger}_L {\rm{diag}}(1,1,1,1) U^{\nu}_L \nu^{\prime}_{L} 
	+ \bar{e}^{\prime}_{Ra} \gamma^\mu (U^{\ell \dagger}_R)_{a4} \tilde{N}_{R}  \right] + h.c.,
\eea
Because of the approximate $SU(2)$ symmetry, the left-handed couplings have $U^{\ell \dagger}_L {\rm{diag}}(1,1,1,1) U^{\nu}_L \simeq 1$. The right-handed interactions are also present with a suppression factor $U^{\ell \dagger}_R$. On the other hand, the quark sector has a somewhat different situation due to the CKM structure. The model parameters $Y_{d_i\tilde D_R}, v_{\Phi}$ and $m_Q$, which decide $U^{u}_{L,R}$ and  $U^{d}_{L,R}$, should be chosen to realize the measured CKM values. In our analysis, we use moderate values of the model parameters which satisfy the CKM values within error bars.

\begin{figure}
	\begin{center}
		\includegraphics[width=0.7 \textwidth]{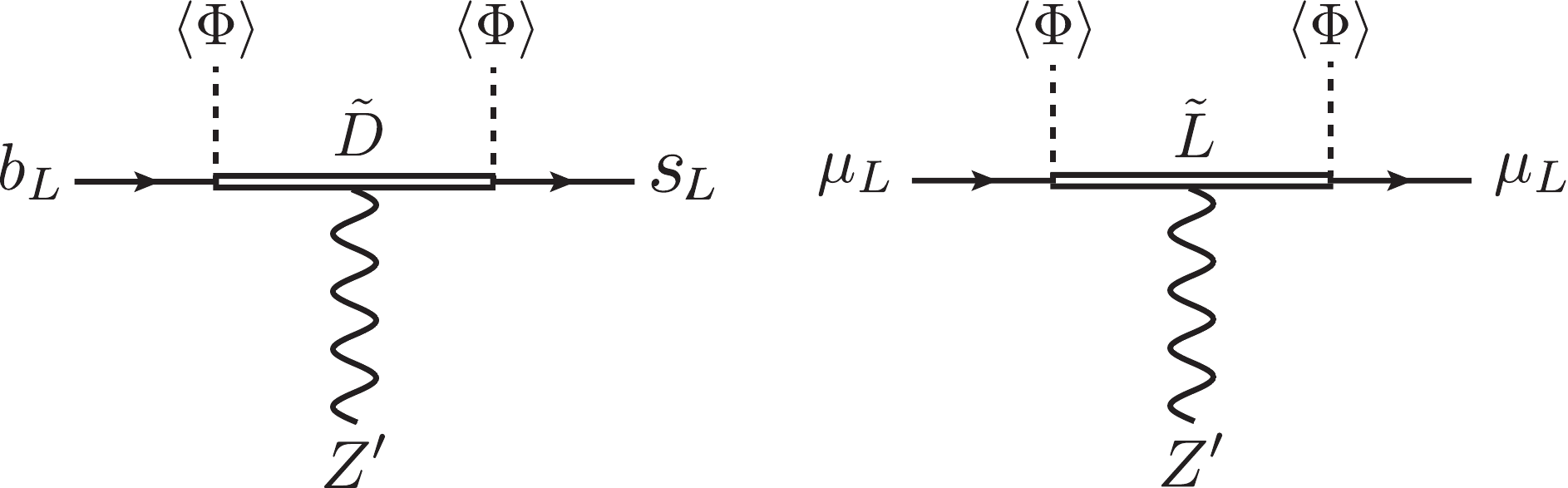}
	\end{center}
	\caption{Examples of the left-handed $Z^{\prime}$ couplings induced by mixing between the vector-like and the SM fermions with mass insertion notation. 
	Here the double line represents the vector-like fermions. 
		\label{fig:eff_Zp}}
\end{figure}

One of our main interests is the $Z'$ couplings, which are given by
\bea
	{\mathcal L}_{Z^{\prime}} &\supset& -g' Z'_\mu \left[\bar{e}^{\prime}_{L} \gamma^\mu U^{\ell \dagger}_L {\rm{diag}}(0,0,0,1) U^{\ell}_L e^{\prime}_{L} +\bar{d}^{\prime}_{L} \gamma^\mu U^{d \dagger}_L {\rm{diag}}(0,0,0,1) U^{d}_L d^{\prime}_{L}  \right].	
\eea
These interactions can be described by integrating out the heavy fermions 
\bea
	\left(U^{\ell \dagger}_L {\rm{diag}}(0,0,0,1) U^{\ell}_L\right)_{ij}\simeq & \frac{v^2_{\Phi}}{2m^2_{L}}Y_{l_i\tilde L_R}Y^*_{l_j\tilde L_R},\\
	\left(U^{d \dagger}_L {\rm{diag}}(0,0,0,1) U^{d}_L\right)_{ij}\simeq & \frac{v^2_{\Phi}}{2m^2_{Q}}Y_{d_i\tilde L_R}Y^*_{d_j\tilde L_R}.
\eea
and this situation is schematically drawn in Fig. \ref{fig:eff_Zp}. 
If we focus on only relevant Yukawa couplings $Y_{b\tilde D_R},~Y_{s\tilde D_R}$ and $Y_{\mu\tilde L_R}$ to the $R_K$ anomaly, they yield  
\begin{align}
{\cal L}_{Z^{\prime}} \supset &-Z^{\prime}_{\mu}\bigg(g^L_{sb}~\bar{s}_L\gamma^{\mu}b_L+g^L_{\mu\mu}~\bar{\mu}_L\gamma^{\mu}\mu_L+g^L_{bs}~\bar{b}_L\gamma^{\mu}s_L+g^L_{ss}~\bar{s}_L\gamma^{\mu}s_L+g^L_{bb}~\bar{b}_L\gamma^{\mu}b_L
 \bigg), \label{eff_Zp}
\end{align} 
where 
\begin{align}
g^L_{d_id_j}=g^{\prime}\frac{v^2_{\Phi}}{2m^2_Q}Y_{d_i\tilde D_R}Y^*_{d_j\tilde D_R},
\hspace{1cm}
g^L_{\mu\mu}=g^{\prime}\frac{v^2_{\Phi}}{2m^2_L}Y_{\mu\tilde L_R}Y^*_{\mu\tilde L_R}, \label{eff_Zp_gL}
\end{align}
proportional to the mixing angles,
and we omit the prime in the fermion fields.
The first two terms in Eq. (\ref{eff_Zp}) lead to the $b\to s \mu\mu$ transition, and
the three Yukawa couplings also produce the $Z^{\prime}$ couplings to the up quark and neutrino sectors, such as $\bar{t}tZ^{\prime}$ and $\bar{\nu}_{\mu}\nu_{\mu}Z^{\prime}$, due to the $SU(2)$ symmetry. Note that the right-handed $Z^{\prime}$ couplings are also present, however, they are induced by the inert scalar loop and are numerically suppressed. Therefore, our current study does not take them into account.

\begin{figure}
	\begin{center}
		\includegraphics[width=0.7 \textwidth]{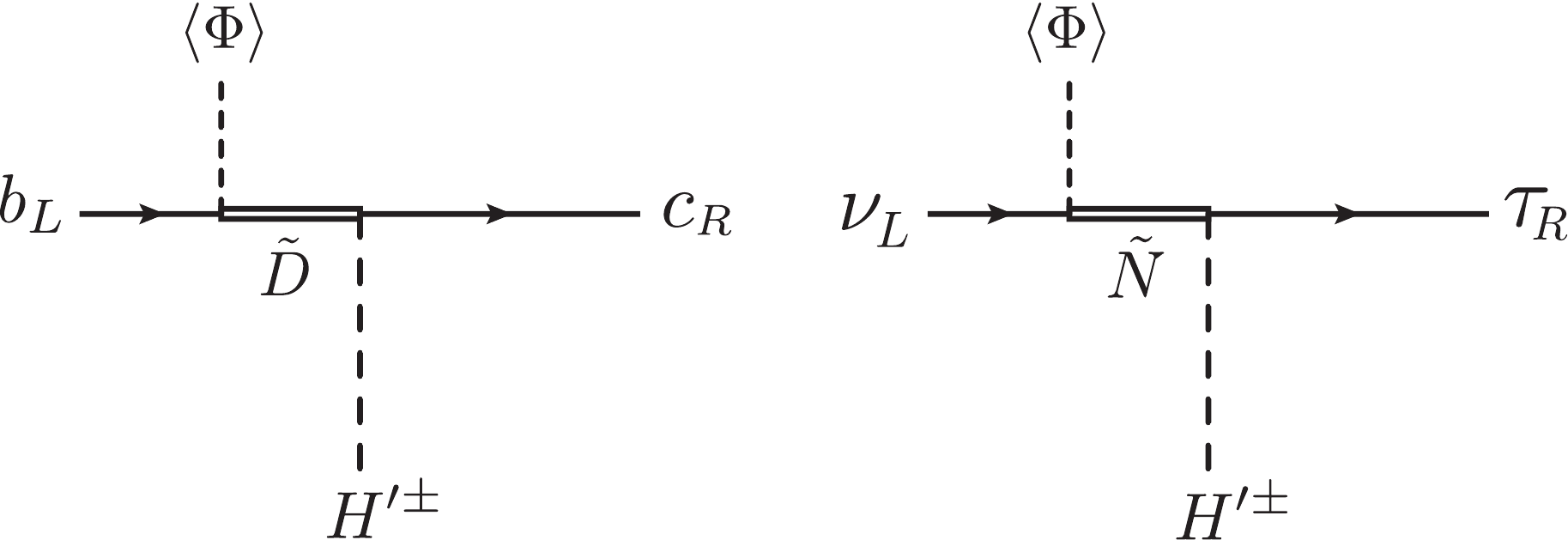}
	\end{center}
	\caption{Examples of the charged scalar couplings through mixing between the vector-like and the SM fermions  with mass insertion notation. 
	Here the double line represents the vector-like fermions. 
		\label{fig:eff_Hp}}
\end{figure}

The other intriguing interactions are the Yukawa couplings between the inert-Higgs doublet and the SM fermions, in particular, the charged Higgs couplings $\bar{c}bH^+$ and $\bar{\nu}\tau H^+$. Using mass-insertion method as in the $Z^{\prime}$ coupling, we obtain the Yukawa couplings
\begin{align}
{\cal L}_{H^{\prime}}\supset&-\left({\cal Y}_{c_{R}b_{L}}\bar{c}_Rb_L+{\cal Y}_{c_{L}b_{R}}\bar{c}_Lb_R\right)H^{\prime +}-{\cal Y}_{\nu_{\mu}\tau_{R}}\bar{\nu}_{L\mu}\tau_R H^{\prime +}+{\rm h.c.}, \label{eff_Yukawa}
\end{align}
with
\begin{align}
{\cal Y}_{c_{R}b_{L}}&=-\frac{v_{\Phi}}{\sqrt{2}m_Q}{\rm y}^{\prime *}_{Q_Lc}Y^*_{b\tilde D_R},\hspace{0.5cm}
{\cal Y}_{c_{L}b_{R}}=\frac{v_{\Phi}}{\sqrt{2}m_Q}Y_{c\tilde D_R}{\rm y}^{\prime}_{Q_Lb},\hspace{0.5cm}
{\cal Y}_{\nu_{\mu}\tau_{R}}=\frac{v_{\Phi}}{\sqrt{2}m_L}Y_{\nu_{\mu}\tilde N_R}{y}^{\prime}_{L_L\tau}.
\end{align}
These interactions are originated from the mixing between the vector-like and SM fermions as seen in Fig. \ref{fig:eff_Hp}. Their dependences on $v_{\Phi}/m_{Q,L}$ are different from those in Eq. (\ref{eff_Zp_gL}). 
The neutral components in the inert Higgs also induce the Yukawa couplings as in Eq. (\ref{eff_Yukawa}). In our study, we assume that their masses are degenerate. This assumption produces a simple situation where some scalar operators disappear due to the  degeneracy as mentioned later.

As we will see in the next section, while the explanation of $R_K$ anomaly needs nonzero $g^L_{sb}$ and $g^L_{\mu\mu}$, the $R_D$ anomaly requires nonzero ${\cal Y}_{cb}$ and ${\cal Y}_{\nu\tau}$. From the phenomenological point of view, we consider a minimal setup in which these relevant couplings are exclusively focused on. Simultaneously, it implies nonzero $Y_{b\tilde D_R},~Y_{s\tilde D_R},~Y_{\mu\tilde L_R}$ and ${\rm y}^{\prime}_{Q_Lb},~{\rm y}^{\prime}_{Q_Lc},~{\rm y}^{\prime}_{L_L\tau}$. From now on, we denote them as

\bea
	Y_b \equiv Y_{b\tilde D_R}, \quad Y_s \equiv Y_{s\tilde D_R},  \quad Y_\mu \equiv Y_{\mu\tilde L_R},
\eea
and
\bea
	y'_\tau \equiv {\rm y}'_{L_L\tau}, \quad y'_b \equiv {\rm y}'_{Q_Lb},  \quad y'_c \equiv {\rm y}'_{Q_Lc}.
\eea
We also assume that the Yukawa couplings are real, and a flavor hierarchy $|Y_b|\gg |Y_s|$.

\section{Rare $B$ Decay Anomalies}
\label{sec:banomaly}
The relevant effective Hamiltonians to $b\to s \mu^+\mu^-$ and $b\to c\tau^-\bar{\nu}$   are given by \cite{Altmannshofer:2013foa,Descotes-Genon:2013wba,Beaujean:2013soa} 
\begin{align}
{\cal H}_{\rm eff}(b\to s\mu^+ \mu^-)&=-\frac{4G_F}{\sqrt{2}}V_{tb}V^*_{ts}\frac{e^2}{16\pi^2}\bigg[C_9(\bar{s}\gamma^{\mu}P_Lb)(\bar{\mu}\gamma_{\mu}\mu)+C_{10}(\bar{s}\gamma^{\mu}P_Lb)(\bar{\mu}\gamma_{\mu}\gamma_5\mu) \bigg],\\
{\cal H}_{\rm eff}(b\to c\tau^-\bar{\nu})&=\frac{4G_F}{\sqrt{2}}V_{cb}\bigg[C_Sm_b(\bar{c}P_Rb)(\bar{\tau}P_L\nu)+C_S^{\prime}m_b(\bar{c}P_Lb)(\bar{\tau}P_L\nu) \bigg].
\end{align}
The Wilson coefficients of $C_9 $ and $C_{10}$ are induced by the $Z^{\prime}$ interactions in Fig. \ref{fig:eff_Zp}. Integrating out the $Z'$ boson, the Wilson coefficients are given by
\begin{align}
C_9=-C_{10}&=-\frac{1}{2m^2_{Z'} C^{bs}_{\rm SM}}g^L_{sb}g^L_{\mu\mu} 
=-\frac{v^2_{\Phi}}{8C^{bs}_{\rm SM}m^2_Qm^2_L}Y_{s}Y_{b}Y_{\mu}^2, \label{C9_10}
\end{align}
with $C^{bs}_{\rm SM}=\frac{4G_F}{\sqrt{2}}V_{tb}V^*_{ts}\frac{e^2}{16\pi^2}$. The final expression in Eq.~(\ref{C9_10}) is obtained by using the couplings defined in Eq. (\ref{eff_Zp_gL}). Although $m_{Z^{\prime}}$ dependence appears in the middle of the Eq.~(\ref{C9_10}), it is canceled out by those in $g^L_{sb}$ and $g^L_{\mu\mu}$. Thus, $v_{\Phi}$ is finally left only in the numerator. 
The nonzero Yukawa, $Y_b,~Y_s$ and $Y_{\mu}$, lead to $Y_q=(V_{\rm CKM})_{qs}Y_s+(V_{\rm CKM})_{qb}Y_b$ for $q=u,c,t$ and $Y_{\nu_i}=(V_{\rm PMNS})_{\nu_i \mu}Y_{\mu}$ for $i=e,\mu,\tau$.

For the $R_D$ anomaly, neglecting the neutrino flavor textures which is irrelevant to our study,  we simply assume that the PMNS matrix is a unit matrix, which results in $Y_{\nu_{\mu}}=Y_{\mu}$. The Yukawa coupling allows the flavor-violating charged Higgs coupling $\bar{\tau}P_L\nu_{\mu}H^-$. As seen in \cite{Bhattacharya:2016mcc}, such a new flavor-violating process would also contribute to the $R_D$ anomaly. Therefore, the charged scalar interactions in Fig. \ref{fig:eff_Hp} produce
\begin{align}
C_S^{\prime}&=-\frac{1}{C^{bc}_{\rm SM}m^2_{H^{\prime}}m_b}{\cal Y}_{c_Rb_L}{\cal Y}^*_{\nu_{\mu }\tau_R}
=\frac{v^2_{\Phi}}{2C^{bc}_{\rm SM}m^2_{H^{\prime}}m_bm_Qm_L}{y}^{\prime }_{c}{y^{\prime }}_{\tau}Y_{b}Y_{\mu },\label{Csp}\\
C_S&=-\frac{1}{C^{bc}_{\rm SM}m^2_{H^{\prime}}m_b}{\cal Y}_{c_Lb_R}{\cal Y}^*_{\nu_{\mu }\tau_R}=-\frac{v^2_{\Phi}}{2C^{bc}_{\rm SM}m^2_{H^{\prime}}m_bm_Qm_L}{y}^{\prime}_{b}{y}^{\prime}_{\tau}Y_{c}Y_{\mu}, \label{Cs}
\end{align}
with  $C^{bc}_{\rm SM}=\frac{4G_F}{\sqrt{2}}V_{cb}$. The third terms are written by Eq. (\ref{eff_Yukawa}).  It is seen that the model parameters appearing in $C_9$ and $C_{10}$ in the Eq. ({\ref {C9_10}}) are correlated to $C_S^{\prime}$ and $C_S$ except $m_{H^{\prime}}$ and $y^{\prime}_{b,c,\tau}$.
Therefore, in our model, the $R_D$ anomaly is related to the $R_K$ anomaly through the common parameters.

In order to determine the possible sizes of these Wilson coefficients, we use flavio 0.21 \cite{flavio}:

\begin{itemize}
\item{$C_9$ and $C_{10}$}

The observables and the corresponding experimental measurements listed in the Appendix in Ref.~\cite{Altmannshofer:2017fio} are used, and Table~\ref{tab:rk} in the last page summarizes them.  We take into account all known correlations 
among observables and approximate the uncertainties as Gaussian.  From Eq.~(\ref{C9_10}), our model holds the relation $C_9=-C_{10}$, therefore essentially only one Wilson coefficient needs to be fitted. Assuming all the UV parameters are real, we define the following real parameters in the global fit:
\begin{eqnarray}
C_{9,\rm new} = C_9 \cdot \exp({i\arg{(C^{bs}_{\rm SM})}}) = -C_{10,\rm new} = -C_{10} \cdot \exp({i\arg(C^{bs}_{\rm SM})})  \label{C9_new}
\end{eqnarray}
The Bayesian method is employed in the global fit with following procedures: we first obtain the likelihood function with a single argument $C_{9,\rm new}$ from flavio FastFit class, and assume a uniform prior probability of $C_{9,\rm new}$ ranges from -3 to 3. 
We then use pymultinest~\cite{Buchner:2014nha} to implement a Monte Carlo sampling, and finally obtain the posterior probability shown in Fig.~\ref{fig:c9fit} with corresponding 1$\sigma$ and 2$\sigma$ uncertainties, which is plotted using Superplot~\cite{Fowlie:2016hew}. In the figure, the best-fit point is indicated by a star, and $1\sigma$ and 2$\sigma$ regions are represented by blue and green lines on the top of curve. The obtained values are in agreement with the current observed branching ratio for $B^{0,\pm}\to K^{*0,\pm} \mu\mu,~B_s\to \phi \mu\mu,~B\to X_s\mu\mu$ and $B_s\to\mu\mu$.

\begin{figure}
	\begin{center}
		\includegraphics[width=0.4 \textwidth]{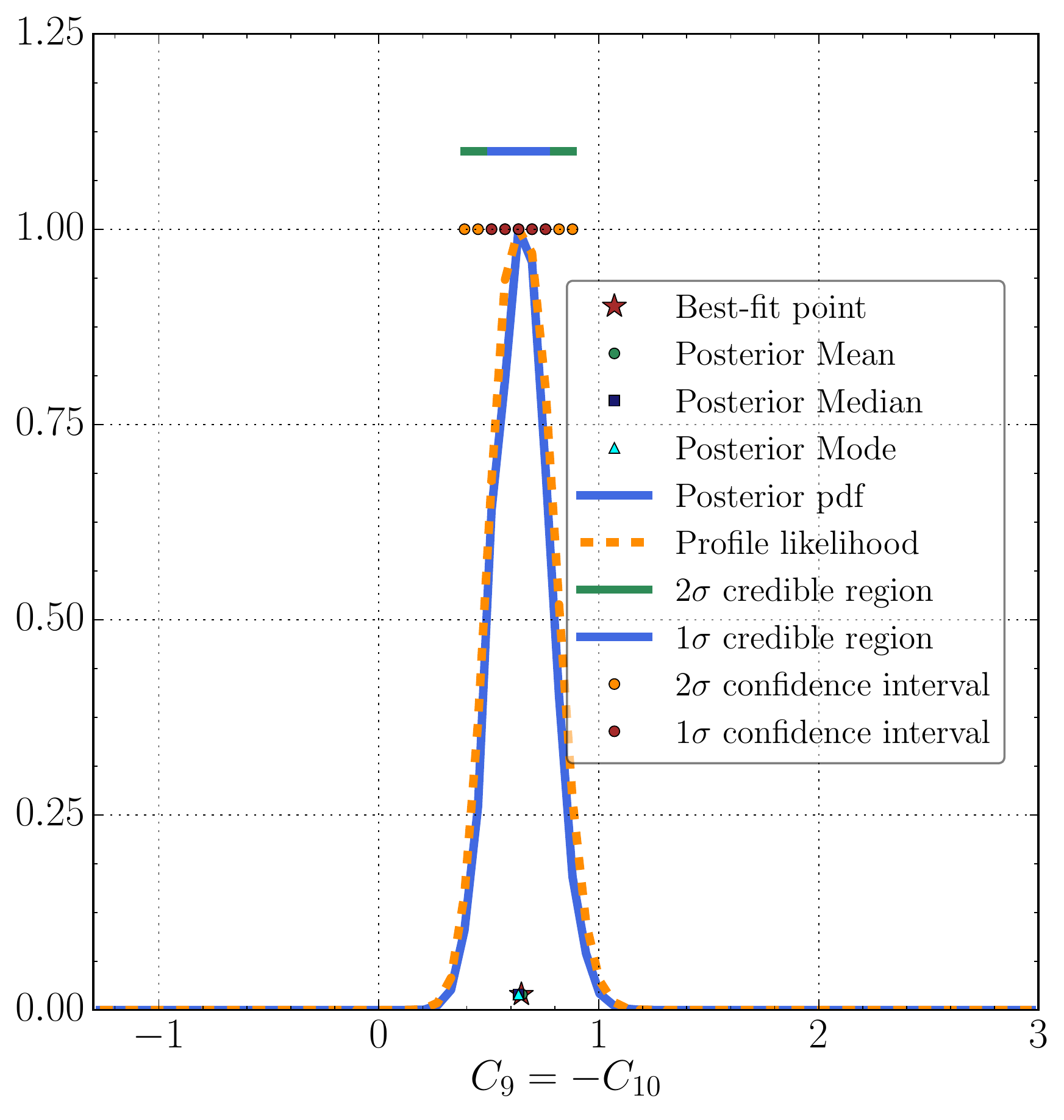}
	\end{center}
	\caption{The likelihood as function of the $C_{9,\rm new}$ from the Bayesian fit. The blue and green lines on the top of curve represent the $1\sigma$ and $2\sigma$ regions. The star at the bottom is the best-fit point.
		\label{fig:c9fit}}
\end{figure}

\item{$C_S$ and $C_S^{\prime}$}

For the $R_{D^*}$, measurements in Ref.~\cite{Sato:2016svk,Abdesselam:2016xqt,Aaij:2015yra} are used, while those in Ref.~\cite{Lees:2013uzd,Huschle:2015rga} are taken into account for the $R_D$ . We use FastFit in flavio to obtain the 2-D global fit for $C_S$ and $C'_S$, which is shown in Fig.~\ref{fig:cscsp}. The four blue regions are $1\sigma,~2\sigma$ and 3$\sigma$ allowed regions by fitting with the $R_D$ and $R_{D^*}$ measurements mentioned above. The light and dark red region are excluded by {{the current LHC limit ${\rm BR}(B^-_c\to\tau^-\bar{\nu})<30\%$~\cite{Alonso:2016oyd}, and recasted LEP limit}} ${\rm BR}(B^-_c\to\tau^-\bar{\nu})<10\%$~\cite{Akeroyd:2017mhr}, respectively.  
We find that among four favored regions, only two of them are more preferred after considering the constraints from the ${\rm BR}(B^-_c\to\tau^-\bar{\nu})$ limits.

\begin{figure}
	\begin{center}
		\includegraphics[width=0.5\textwidth]{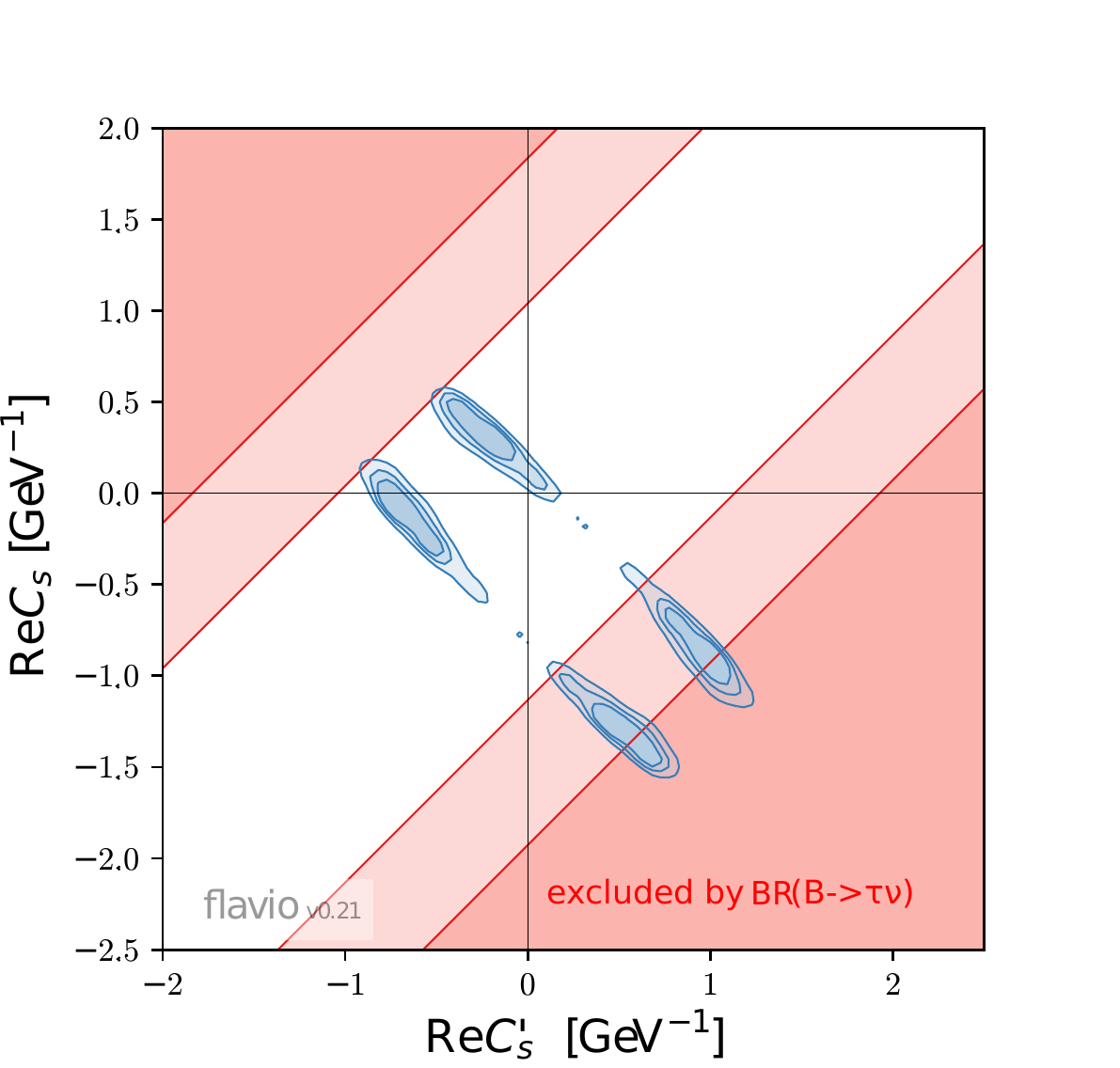}
	\end{center}
	\caption{The two dimensional contour for $C_S$ and $C'_S$ as a result of the FastFit. The blue regions corresponds to 1$\sigma$, 2$\sigma$ and 3$\sigma$ regions in agreement with $R_D$ and $R_{D^*}$ measurements. The red and light red regions are excluded {{by the current LHC limit ${\rm BR}(B^-_c\to\tau^-\bar{\nu})<30\%$ \cite{Alonso:2016oyd}, and recasted LEP limit}} ${\rm BR}(B^-_c\to\tau^-\bar{\nu})<10\%$ \cite{Akeroyd:2017mhr}, respectively.
		\label{fig:cscsp}}
\end{figure}

For later use, we project the two dimensional contour to a 1-dimensional fit by setting $C_S=0$. 
The Bayesian fit for $C'_S$ alone is obtained using the same strategy as in the fitting of $C_{9,\rm new}$. For this fit, we choose prior probability as uniformly distribute between $-2$ and $0$, the fitting results is shown in Fig.~\ref{fig:cspfit}. As in Fig.~\ref{fig:c9fit}, the best fit value is represented by the star at the bottom, and $1\sigma$ and $2\sigma$ regions are blue and green lines above curve.

\begin{figure}
	\begin{center}
		\includegraphics[width=0.4\textwidth]{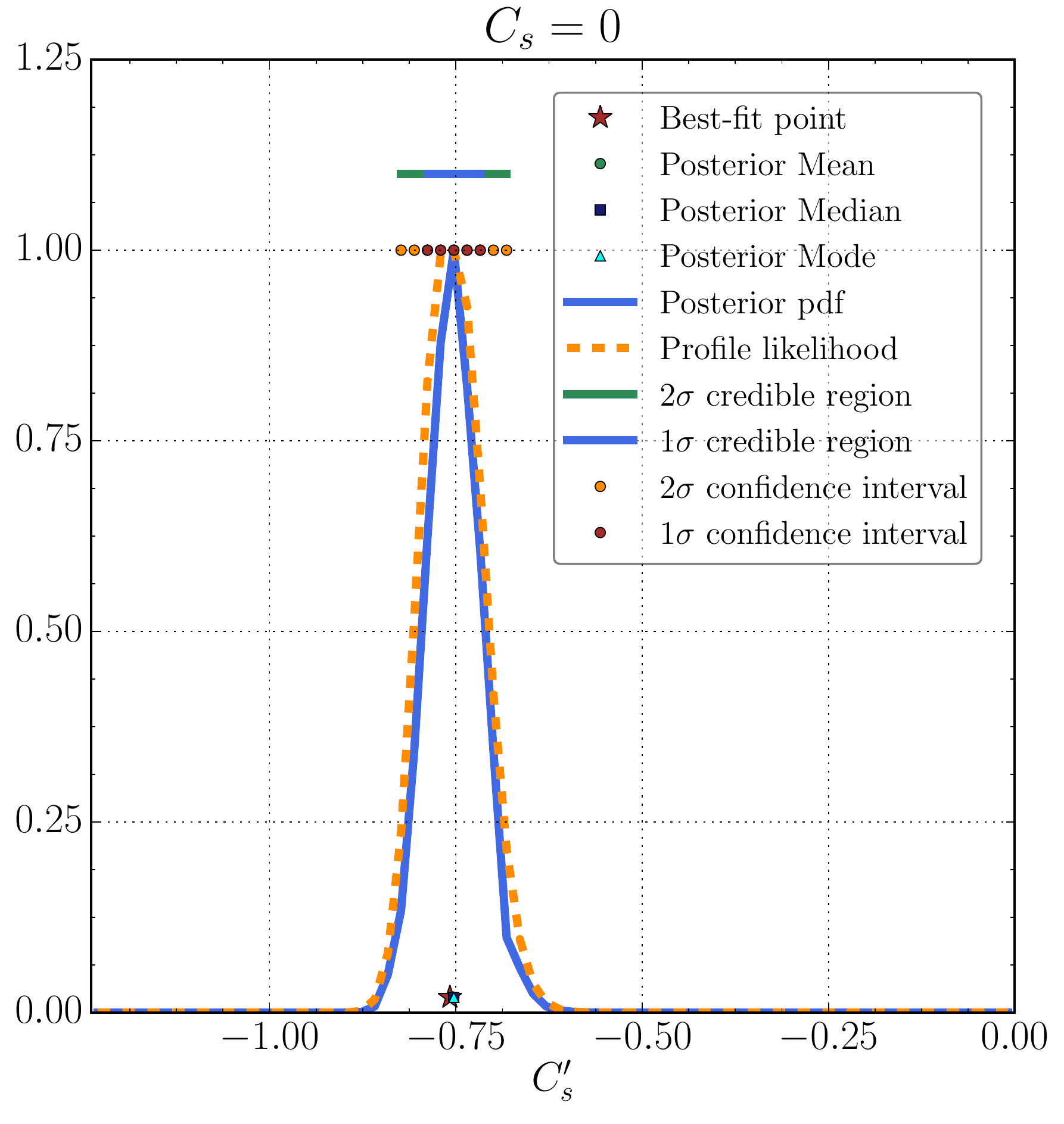}
	\end{center}
	\caption{The likelihood as function of  $C'_{S}$ with $C_S=0$ based on the Bayesian fit. The star indicates the best fit value, and the blue and green lines are $1\sigma$ and $2\sigma$ regions.
		\label{fig:cspfit}
		}
\end{figure}
\end{itemize}

Finally, we summarize the obtained numerical values in Table. \ref{table:general_analysis} and \ref{table:model_analysis}, where
those of $C_S$ and $C_S^{\prime}$ without the constraint from $B^-_c\to\tau^-\bar{\nu}$ transition are listed.
%
\renewcommand{\arraystretch}{1.0}
\begin{table}[h]
\begin{tabular}{l|c|c|c}
& $C_9=-C_{10}$ & $C_S$  & $C_S^{\prime} $ \\
\hline
\hline
~$1\sigma$~ & $[-0.742,~-0.643]$ & $[-0.344,~0.074]$  & $[-0.821,~-0.461]$\\
 &  & $[0.180,~0.517]$  & $[-0.444,~-0.068]$\\
 &  & $[-1.045,~-0.627]$  & $[0.755,~1.109]$\\
  &  & $[-1.498,~-1.153]$  & $[0.351,~0.734]$\\
\hline
$2\sigma$ &  $[-0.767,~-0.618]$ & $[-0.389,~0.128]$ & $[-0.863,~-0.430]$ \\
 &   & $[-0.123,~0.547]$ & $[-0.482,~0.100]$ \\
 &  & $[-0.605,~-0.083]$ & $[0.712,~1.162]$ \\
  &  & $[-1.022,~-0.497]$ & $[0.182,~0.770]$ \\
\hline
$3\sigma$ & $-$ & $[-0.087,~0.684]$ & [-0.919,~-0.234] \\
&  & $[0.462,~1.078]$ & [-0.529,~0.175] \\
&  & $[-0.672,~0.110]$ & [0.520,~1.234] \\
&  & $[-1.045,~-0.431]$ & [0.113,~0.828] \\
\end{tabular}
\caption{Suitable sizes of the Wilson coefficients $C_9,C_{10}$ and $C_S,C_S^{\prime}$ for the $R_K$ and $R_D$ anomalies. Here, the restrictions from $Br(B^-_c\to\tau^-\bar{\nu})$ are not taken into account. }
\label{table:general_analysis}
\end{table}
%
\begin{table}[h]
\begin{tabular}{c|c|c}
  & $C_{9, \rm new}=-C_{10, \rm new}$  & $C_S^{\prime}$ with $C_S=0$ \\
  \hline
  \hline
~$1\sigma$~ & $[0.515,~-0.72]$ & $[-0.788,~-0.717]$  \\
\hline
$2\sigma$  & $[0.433,~0.843]$  & $[-0.824,~-0.682]$
\end{tabular}
\caption{Desired values of $C_{9,\rm new}$ defined in Eq. (\ref{C9_new}) and $C_S^{\prime}$ on the line of $C_S=0$.  }
\label{table:model_analysis}
\end{table}

\section{Experimental Searches and Constraints}
\label{sec:constraints}

The hidden gauge boson $Z'$ encounters direct constraints from collider searches. 
Depending on the $Z'$ mass, $m_{Z'} = g' v_\Phi$, there are different limits on the signal rate of the $Z'$ production. Here, we discuss the following cases: light $Z'$ case ($m_{Z'} < m_Z$) and heavy $Z'$ case ($m_{Z'} > m_Z$).
\bit
\item When the $Z'$ is lighter than the $Z$ mass, some viable parameter regions exist. 
Since the $Z'$ has no coupling to the electron, LEP searches cannot provide 
direct constraint on the light $Z'$. 
Furthermore, the Tevatron~\cite{Aaltonen:2011gp,Abazov:2010ti} and LHC~\cite{Aaboud:2017buh,Khachatryan:2016zqb} searches for $Z'$ to dilepton final state only apply to the case of $m_{Z^{\prime}}>100~$GeV.
The relevant limit to the light $Z'$ case comes from the LHC searches at $ p p \to Z\to 4\mu$. Its typical SM process is through the off-shell $Z$ mediated $2\mu$ decay, while the 
light $Z'$ could be on-shell when {{$m_{Z'} < m_{Z} - 2 m_\mu$}}. 
A detailed analysis on how to recast the current LHC search limit has been done in Ref.~\cite{Altmannshofer:2014pba}. 
Mapping their analyses to our model, we obtain the constraint shown in Fig.~\ref{fig:zprime4l}. In the figure, the orange line indicates the $g^{\prime}$ coupling as a function of $m_{Z^{\prime}}$ with $v_\Phi=700$ GeV. The excluded region corresponds to the blue region. For $v_\Phi=700$ GeV, we find that the region of $m_{Z'}\lesssim 10$~GeV and $50~{\rm GeV}\lesssim m_{Z'}$ still have some spaces for $10^{-3}<g^{\prime}<0.14$. However, one should note that for $m_{Z'}\lesssim 10$~GeV, we are not able to integrate out the $Z'$ particle when deriving the Wilson coefficients $C_9$ and $C_{10}$ in Eq.\ref{C9_10}.

\begin{figure}
	\begin{center}
		\includegraphics[width=0.5\textwidth]{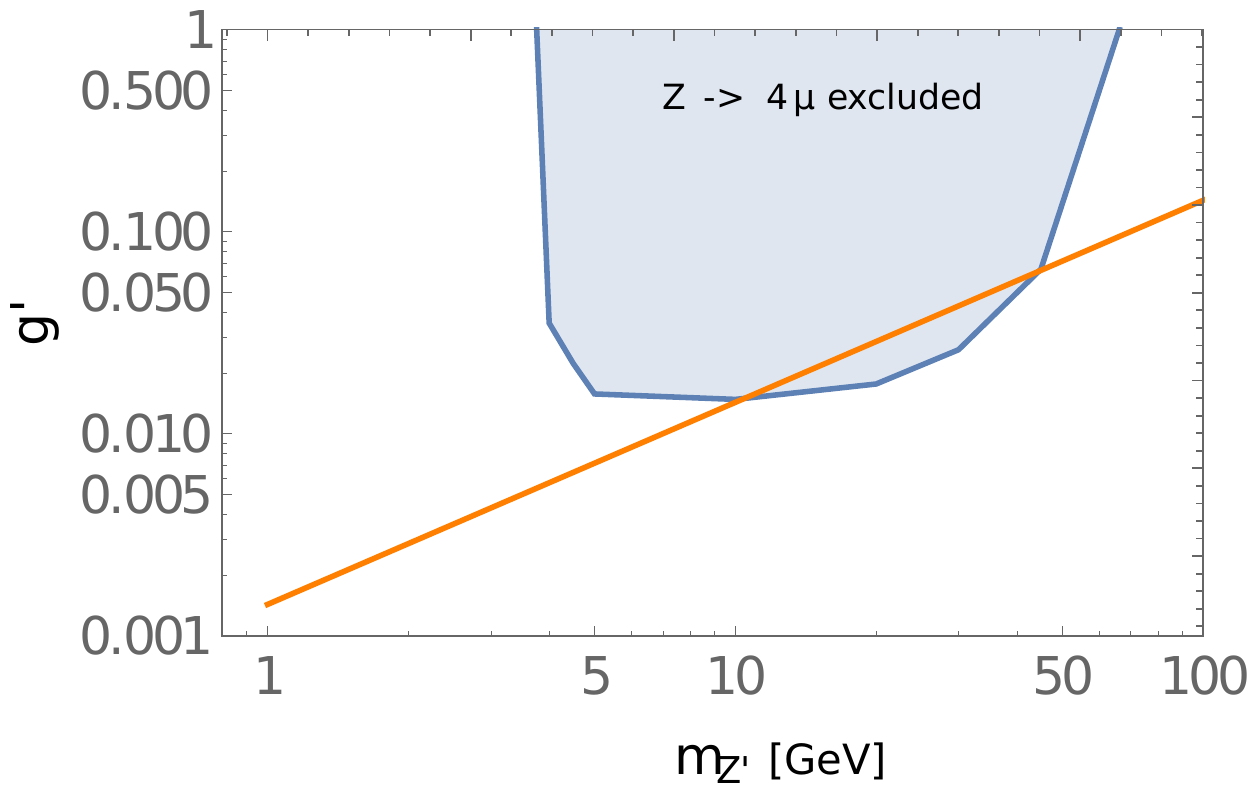}
	\end{center}
	\caption{Constraint on low mass $Z'$ from $Z\to 4\mu$ search at LHC. The blue region is excluded by the current LHC searches. The orange line represents our benchmark model with $v_\Phi=700$ GeV. 
\label{fig:zprime4l}}
\end{figure}

\item In the case of the heavy $Z^{\prime}$, the LHC searches in the di-muon final states put the tightest constraints on its mass. 
The current limit on $m_{Z^{\prime}}$ is around $3\sim4$ TeV ~\cite{Aaboud:2017buh,Khachatryan:2016zqb}. In order to apply this limit to our case, we should first take into account the suppressed coupling to the first generation quarks by the small mixing with vector-like fermions, which results in small production cross section.
Moreover, although the LHC searches assume that the decay branching ratio to the muon or electron lepton is $100\%$,  our $Z'$ can also decay into the SM quarks, 
and other light $U(1)_H$ charged particles. Therefore, the branching ratio could be much smaller than the assumed value. 
We estimate $pp\to Z'$ production cross section with Madgraph~\cite{Alwall:2014hca}, and demonstrate the constraint on $\sigma(pp\to Z'\to \mu\mu)$ in Fig.~\ref{fig:zprime_heavy} with benchmark values: $v_\Phi=700$ GeV, $Y_b=-1$ and $Y_s=0.0184$. In addition to the muons, another decay processes of the bottom, top, neutrino, charged scalar, scalar $(s, H)$ and pseudo-scalar ($A$) are also present. A naive estimation implies that the $Z'$ decay branching ratio to the muons is roughly $1/10.5$. 
The cross section times branching ratio is shown in the red curve. According to the figure, the $Z'$ mass can be as low as $1.5$ TeV. 
Furthermore, the branching ratio can be reduced if additional decay channels come from 
unknown hidden gauge sector particles, which are irrelevant to our flavor study. 
However, if we allow opening additional decay channels, the constraint could be significantly relaxed. 
For example, taking the branching ratio to be ${\rm BR}(Z^{\prime}\to\mu\mu)=1/20$, which is shown in the green curve, the $Z'$ mass can be lower than 750 GeV.

\eit

\begin{figure}
	\begin{center}
		\includegraphics[width=0.5\textwidth]{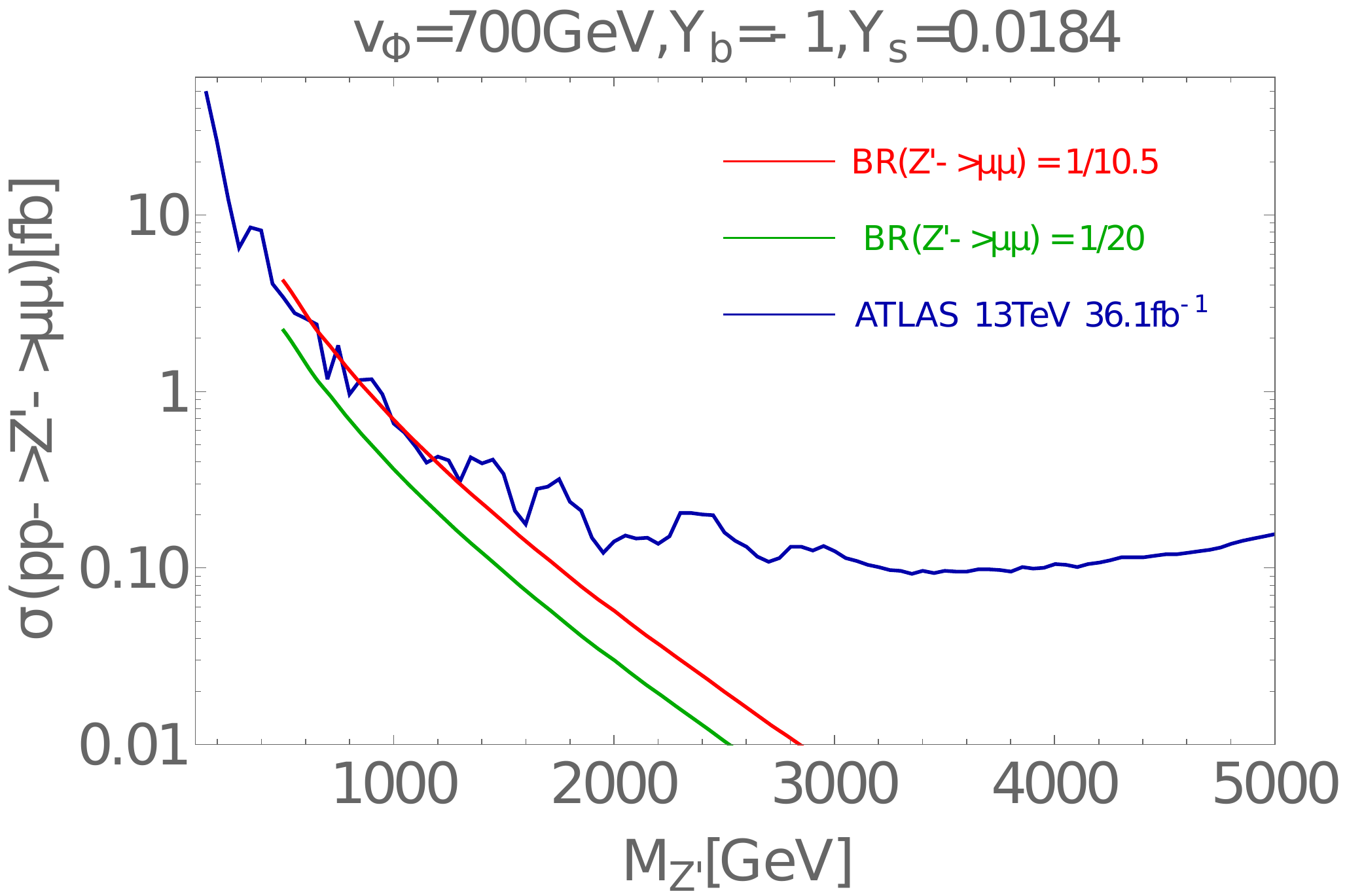}
	\end{center}
	\caption{Constraint on $Z'$ with $v_\Phi=$ 700 GeV, $Y_b=-1$ and $Y_s=0.0184$. The blue line is the $95\%$ C.L exclusion limit by ATLAS experiments~\cite{Aaboud:2017buh}. The red and green curves correspond to the branching ratio of ${\rm BR}(Z'\to \mu\mu)=1/10.5$ and ${\rm BR}(Z'\to \mu\mu)=1/20$ (including new hidden particles), respectively.
\label{fig:zprime_heavy}}
\end{figure}

Next, we briefly comment on the constraints on the mediators. In \cite{Aaboud:2017zfn}, the vector-like quark search is discussed with its decay into $Wb$, and the current limit  is $m_{Q} > 1.1$ TeV. On the other hand, studies about the multi-lepton final state~\cite{Dermisek:2014qca} show that the limit on the vector-like lepton is $m_{L} > 500$ GeV. In our numerical analysis, we take the benchmark point $m_{Q}=m_{L}=1.4$ TeV, which is consistent with these limits.
Regarding the inert-Higgs doublet, the charged scalar receives the tightest bounds.  
The LHC searches for $H^\pm$ are all associated with the top quark production through $tbH^\pm$ coupling. Therefore, we can evade any constraints from the LHC by setting $y'_t=y'_b=0$ such that the $tbH^\pm$ coupling vanishes, and it follows that $C_S=0$. Actually, we investigate the scenario that $y'_t$ and $y'_b$ are nonzero and calculate the production cross section for the process $pp\to \bar{t} bH^+$ with Madgraph for a benchmark point: $y'_t=y'_b=1$, $m_Q=1.4$ TeV, $v_\Phi=700$ GeV, $Y_b=-1$ and $Y_s=0.0184$. It is found that $m_{H^\pm}>300$ GeV is consistent with the ATLAS experiments~\cite{Aad:2015typ,Aaboud:2016dig} provided that  ${\rm BR}(H^+\to\tau\nu)<80$\%. Again,  if taking into account the $H^{\prime\pm}$ decay channel, $H^{\prime +}\to t\bar{b}$, $H^+\to c\bar{b}$, $H^{\prime+}\to c\bar{s}$ and $H^{\prime +}\to \tau\nu$, we find that the requirement ${\rm BR}(H^+\to\tau\nu)<80$\% is easily satisfied.

In the following, we consider various constraints from flavor physics and low energy observables: 
\begin{itemize}

\item{$B_s-\bar{B_s}$ mixing}

\begin{figure}
	\begin{center}
		\includegraphics[width=0.4\textwidth]{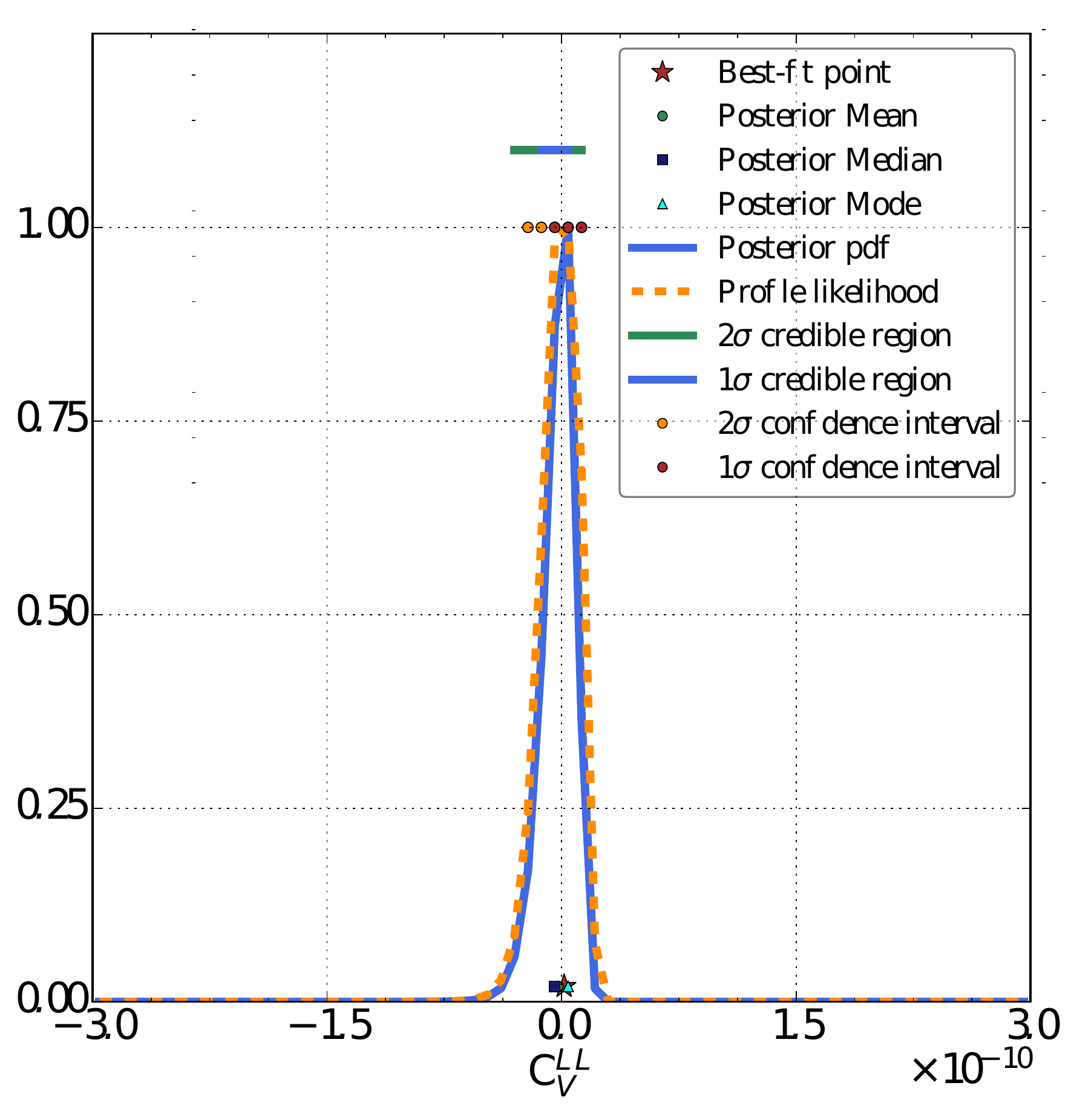}
	\end{center}
	\caption{The normalized likelihood as function of the $C_V^{LL}$ coefficient using the Bayesian fit.
		\label{fig:cvllfit}}
\end{figure}

The $\bar{s}bZ^{\prime}$ coupling induces a relevant operator to the $B_s-\bar{B_s}$ mixing
\begin{align}
{\cal H}_{\rm eff}=-C_V^{LL}(\bar{s}_L\gamma^{\mu}b_L)(\bar{s}_L\gamma_{\mu}b_L)
\end{align}
with $C^{LL}_V=-\left(g^L_{sb} \right)^2/m^2_{Z^{\prime}}$. We calculate the allowed region of this operator using the measurements of $\Delta M_s$ for $B_s-\bar{B}_s$ mixing in Ref.~\cite{Amhis:2014hma}. In our analysis, the Bayesian fit is also employed. The likelihood function is approximated by Gaussian probability distribution function with mean as difference of the theoretical prediction and experimental central value, and the variance as the the quadratic sum of the theoretical uncertainty and the experimental uncertainty. The prior probability distribution is taken as uniform distribution between $-3\times 10^{-10}$ to $3\times 10^{-10}$. The fitting results are shown in Fig.~\ref{fig:cvllfit}, and it is found that $|C_V^{LL}| < 1.29\times 10^{-11}$ at 1~$\sigma$. Note that, although scalar operators ${\cal O}^{LL}_S$ and ${\cal O}^{RR}_S$ listed in \cite{Buras:2012fs} are caused by neutral and CP-odd scalars in the inert-Higgs doublets, they vanish as long as the two scalar masses are degenerate. Also, we do not consider an operator ${\cal O}^{LR}_S$ since its Wilson coefficient is proportional to $y^{\prime}_s$ that is irrelevant to our current study.

\item{Neutrino trident production}

Neutrino trident production gives an upper limit on $g^L_{\mu\mu}/m_{Z^{\prime}}=v_{\Phi}Y^2_{\mu L_R}/2m^2_L$. Its cross section is given by \cite{Altmannshofer:2014pba, Alok:2017jgr, Alok:2017sui}
\begin{align}
\frac{\sigma_{\rm SM+NP}}{\sigma_{\rm SM}}=\frac{1}{1+(1+4s^2_W)^2}\left[\left\{1+v^2\left(\frac{g^L_{\mu\mu}}{m_{Z^{\prime}}}\right)^2\right\}^2+\left\{1 +4s^2_W+v^2\left(\frac{g^L_{\mu\mu}}{m^2_{Z^{\prime}}}\right)^2 \right\}^2 \right]. \label{NT}
\end{align}
The experimental measurement~\cite{Mishra:1991bv} shows ${\sigma_{\rm exp}}/{\sigma_{\rm SM}}=0.82\pm0.28$. For $m_L=1.4~{\rm TeV}$ and $v_{\Phi}=700~{\rm GeV}$, the $1\sigma$ region gives an upper bound on the Yukawa coupling $|Y_{\mu L_R}|\lesssim2.1$.

\begin{figure}
	\begin{center}
		\includegraphics[width=0.4\textwidth]{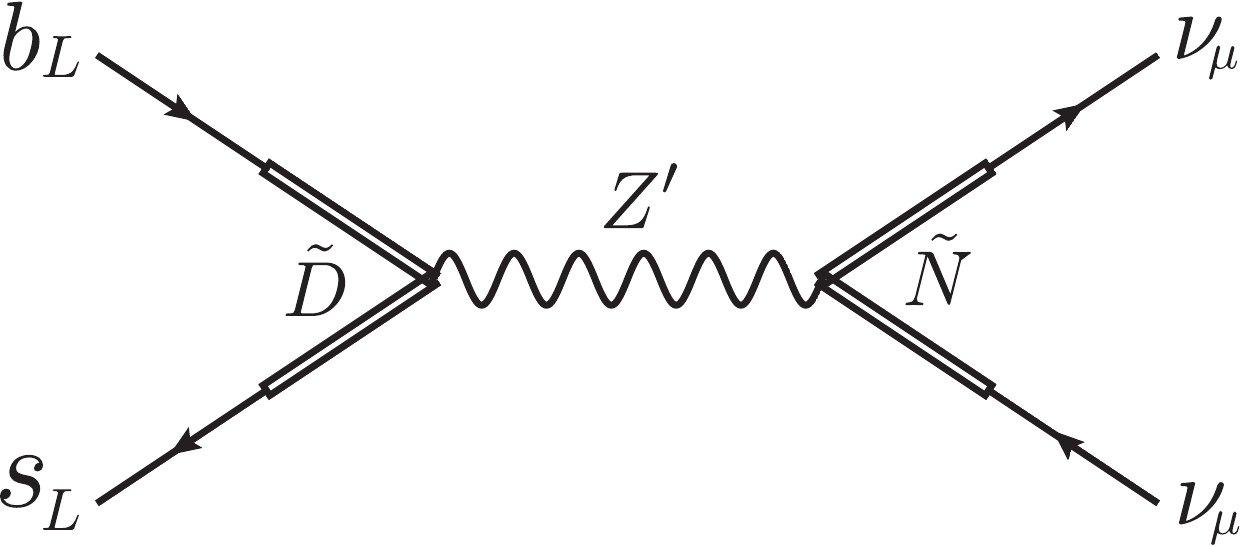}
	\end{center}
	\caption{The $b\to s \nu\nu$ transition induced by the $Z^{\prime}$ exchanges. 
	Here the double line represents the vector-like fermions. 	\label{fig:bsnunu}}
\end{figure}

\item{$B\to K \nu\bar{\nu}$}

In the same way as $b\to s \mu\mu$ process, $b\to s \nu\bar{\nu}$ transition is also induced as shown in Fig.~\ref{fig:bsnunu}. The effective Hamiltonian is \cite{Bhattacharya:2016mcc}
\begin{align}
{\cal H}_{\rm eff}(b\to s\nu_i\bar{\nu}_j)=-\frac{\alpha G_F}{\sqrt{2}\pi}V_{tb}V^*_{ts}C^{ij}_L(\bar{c}\gamma^{\mu} P_Lb)(\bar{\nu}_i\gamma_{\mu}(1-\gamma_5)\nu_j).
\end{align}
Our current setup produces only the 2nd generation neutrino at final states, therefore the Wilson coefficient is described by
\begin{align}
C^{22}_L=-\frac{1}{\sqrt{2}m^2_{Z^{\prime}}}\frac{\pi}{\alpha G_FV_{ts}V^*_{ts}}g^L_{sb}g^L_{\nu_{\mu}\nu_{\mu}},
\end{align}
with $g^L_{\nu_{\mu}\nu_{\mu}}=g^L_{\mu\mu}$.
Following the bound on the Wilson coefficient obtained in \cite{Bhattacharya:2016mcc}, we find that $-13{\rm Re} [C^{22}_L]+|C^{22}_L|^2 \leq 473$.

\begin{figure}
	\begin{center}
		\includegraphics[width=0.7\textwidth]{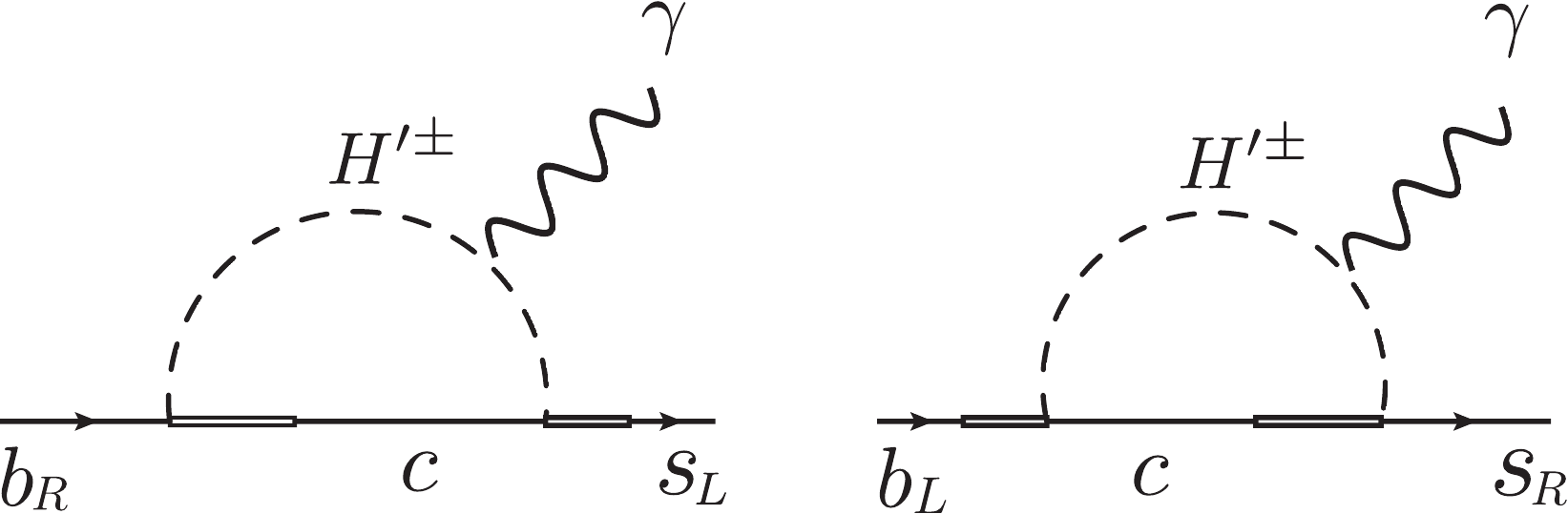}
	\end{center}
	\caption{$b\to s \gamma$ process through the charged scalar loop. The double lines represent the vector-like fermions $\tilde U$ or $\tilde D$. The left diagram is proportional to $y_b^{\prime}y_c^{\prime}$, while the right diagram contains $y_c^{\prime}y_s^{\prime}$.	\label{fig:bsgamma}}
\end{figure}

\item{$b\to s\gamma$}

The Yukawa couplings in Eq. (\ref{eff_Yukawa}) contribute to $b\to s\gamma$ process through the charged scalar loop. In Fig. \ref{fig:bsgamma}, two possible diagrams are drawn. It is seen that while the left diagram contains the product $y^{\prime}_by^{\prime}_c$ related to $C_S$ and $C_S^{\prime}$, the diagram on the right is proportional to $y^{\prime}_cy^{\prime}_s$. Therefore, we include only the left diagram which contains two relevant parameters to $b\to c$ transition. The effective operators are given by \cite{Borzumati:1998tg}
\bea
{\cal H}_{\rm eff}=-\frac{4G_F}{\sqrt{2}}V_{tb}V^*_{ts}\bigg(C_7{\cal O}_7+C_8{\cal O}_8 \bigg)
\eea
with
\bea
{\cal O}_7&=&\frac{e}{16\pi^2}m_b\bar{s}_L\sigma^{\mu\nu}b_RF_{\mu\nu}\\
{\cal O}_8&=&\frac{e}{16\pi^2}m_b\bar{s}_LT^a\sigma^{\mu\nu}b_RG^A_{\mu\nu},
\eea
The Wilson coefficients in our model are 
\bea
C_7&=&\frac{v^2}{V_{tb}V_{ts}^*}{\cal Y}_{c_Lb_R}{\cal Y}^*_{c_Rs_L}\left( \frac{C^0_{7,XY}(x_c)}{m_bm_c}+\frac{C^0_{7,YY}(x_c)}{m^2_c}\right),\\
C_8&=&\frac{v^2}{V_{tb}V_{ts}^*}{\cal Y}_{c_Lb_R}{\cal Y}^*_{c_Rs_L}\left( \frac{C^0_{8,XY}(x_c)}{m_bm_c}+\frac{C^0_{8,YY}(x_c)}{m^2_c}\right),
\eea
where $x_c=m^2_c/m^2_{H^{\prime}}$ and the loop functions $C^0_{7,8}$ are listed in \cite{Borzumati:1998tg}. The Wilson coefficients $C_7$ and $C_8$ are proportional to a product of $Y_sY_cy_b^{\prime}y_c^{\prime}$. While $Y_s$ and $Y_c$ are predicted by the explanation of the $R_{K}$ anomaly, $y_b^{\prime}$ and $y^{\prime}_c$ are related to $C^{\prime}_S$ and $C_S$ in Eqs. ({\ref {Csp}}) and ({\ref {Cs}}). 
According to a global fit of the Wilson coefficients in the $b\to s$ transition in \cite{Descotes-Genon:2015uva}, contributions from new physics to $C_7(\mu_b=5~{\rm GeV})$ are allowed between $-0.04\leq C_7(\mu_b)\leq 0.0$. Taking that the charged scalar mass is $300$ GeV, we find that the loop functions are roughly $O(10^{-5} \sim 10^{-6})$. In addition, the Wilson coefficients receive suppression proportional to $Y_sY_c\sim10^{-4}$. Later, we will check this constraint numerically. It should be noted that, as in the $B_s-\bar{B}_s$ mixing, the neutral scalars do not contribute to the process due to cancellations between $H^{\prime}$ and $A^{\prime}$ contributions for the degenerate mass spectra in the inert Higgs.

\end{itemize}
In addition to the above constraints, lepton-flavor-violating processes, such as $\tau\to 3\mu$ and $\tau\to\mu\gamma$ can also be induced. However, these processes also need irrelevant parameters to the $B$ anomalies, we do not deal with them here.

Finally, we briefly discuss the hidden particles which are only charged under the $U(1)_H$ group.
Although these hidden particles are not relevant to the $B$ physics observables, it provides additional signatures.  
We assume that there is a WIMP dark matter candidate in the hidden sector, and the dark matter particle $\chi$ 
should annihilate into SM particles during freeze-out at early universe. 
In the non-relativistic approximation, 
the annihilation cross section times relative velocity $\langle \sigma v  \rangle$ can be decomposed as 
\bea
\label{eq:xsecvab}
\langle \sigma v  \rangle  = a + b v^2 + {\cal O}(v^4),
\eea
{{where $a$ and $b$ are the $s$-wave and $p$-wave cross sections. }}
We only take the dominant contributions into account. 
When $m_\chi < m_{Z'}$, the dominant annihilation channel is through the $s$-channel process $\bar\chi \chi \to f\bar f$, where $ f = \mu , \nu_\mu, b, t$:
\bea
	\langle \sigma v  \rangle_{\rm s-channel}(\bar\chi \chi \to f\bar f) 
	= \frac{g'^2 g_{Z'ff}^2}{2\pi} \frac{2 m_\chi^2 + m_f^2}{4 m_\chi^2 - m_{Z'}^2} \sqrt{1 - \frac{m_f^2}{m_\chi^2}},
\eea
where $g_{Z'ff}$ denotes the $Z'$ coupling to the fermion $f$. 
When $m_\chi > m_{Z'}$, additional $t$-channel $\bar\chi \chi \to Z' Z' \to f\bar f f\bar f$ {{appears}}. The annihilation cross section is
\bea
	\langle \sigma v  \rangle_{\rm t-channel}(\bar\chi \chi \to Z' Z') 
	= \frac{g'^4}{4\pi } \frac{m_\chi^2 }{(2 m_\chi^2 - m_{Z'}^2)^2} \left(1 - \frac{m_{Z'}^2}{m_\chi^2} \right)^{3/2}.
\eea 
The thermal relic density is written as
\bea
 \Omega_{\rm DM} h^2 = \frac{m_\chi n_\chi}{\rho_{\rm crit}/h^2}  = \frac{ s_0 h^2}{\sqrt{\frac{\pi }{45}}\rho_{\rm crit}} \frac{1} {M_{\rm{pl}} \sqrt{g_{\rm eff}} {\mathcal I}(x_f) },
\eea
where $s_0$ is the entropy density of the present universe, $\rho_{\rm crit}$ the critical density, 
$n_\chi$ the dark matter number, $h$ the hubble parameter, $g_{\rm eff}$ the effective degree of freedom during freeze-out, and ${\mathcal I}(x_f) = \frac{a + 3 b/x_f}{x_f}$, with $x_f = m_\chi/T_f$ and $T_f$ the temperature during freeze-out. 
There are also constraints from the direct detection measurements. However, we expect the limit is not so tight for the light $Z'$, because the size of the gauge coupling $g'$ is typically smaller than $0.1$.
Furthermore, in our model the dark matter only dominantly couples to the third generation quarks inside nucleons. This kinds of scenarios
have been studied in Ref.~\cite{Sage:2016uxt}. From their study, we find that for the light $Z'$, if $g' \sim 0.05$, the direct detection constraints could be escaped.
We will perform detailed study on the dark matter sector in future.

\section{Results}
\label{sec:results}
\begin{figure}
	\begin{center}
		\includegraphics[width=0.43 \textwidth]{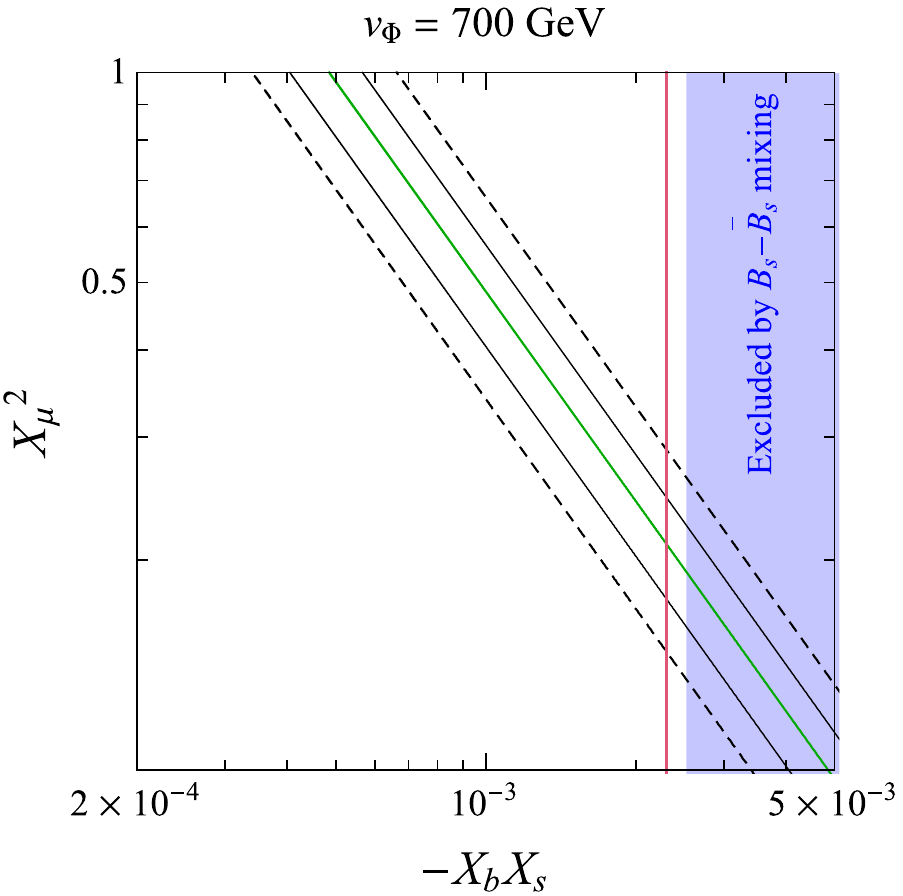}
		\includegraphics[width=0.395 \textwidth]{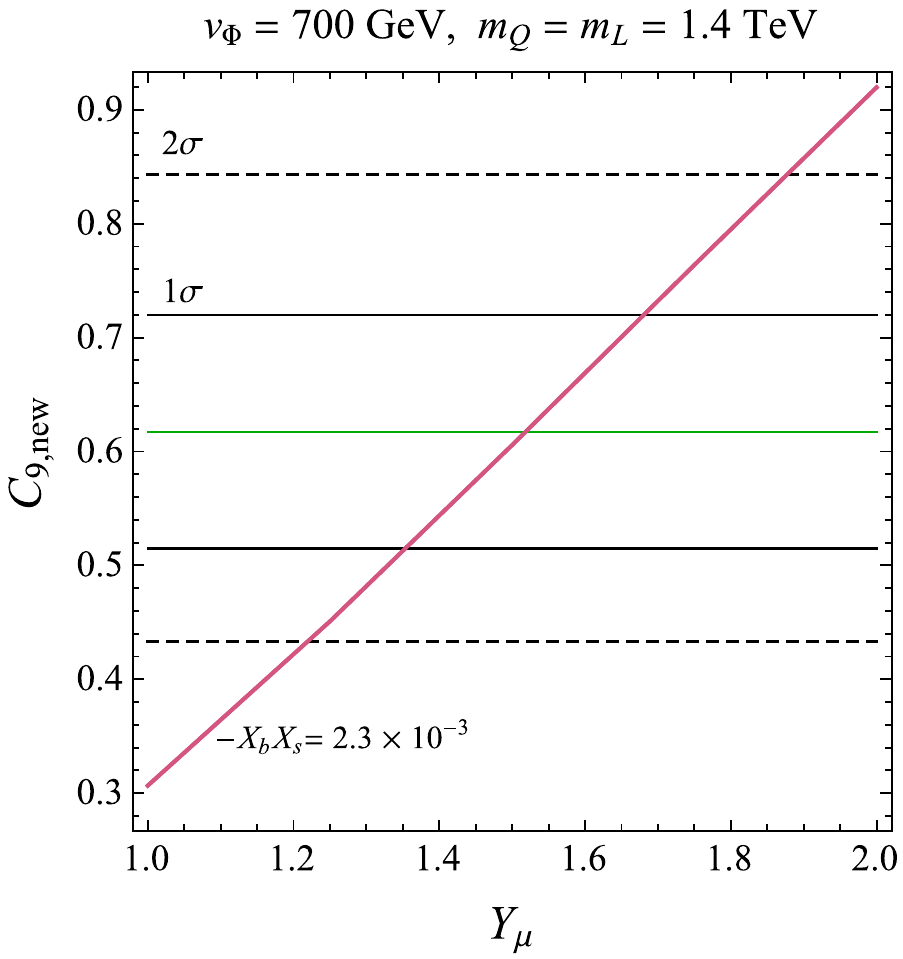}
	\end{center}
	\caption{[Left] $1\sigma$ and $2\sigma$ regions of $C_{9, \rm new}(=-C_{10,\rm new})$ are drawn with black solid and dashed lines, respectively. $v_{\Phi}$ is fixed at $700~$GeV. The best fitted value for the $R_K$ anomaly is indicated by the green line, and the red line represents $-X_bX_s =2.3\times 10^{-3}$. The blue region is excluded by the $B_s-\bar{B_s}$ mixing. [Right] The size of $C_{9, \rm new}$ as a function of $Y_{\mu}$ with the fixed value $-X_bX_s=2.3\times 10^{-3}$. We take  $m_Q=m_L=1.4~$TeV.}
	\label{fig:RK_anomaly} 	
\end{figure}

\begin{figure}[t]
	\begin{center}
		\includegraphics[width=0.4 \textwidth]{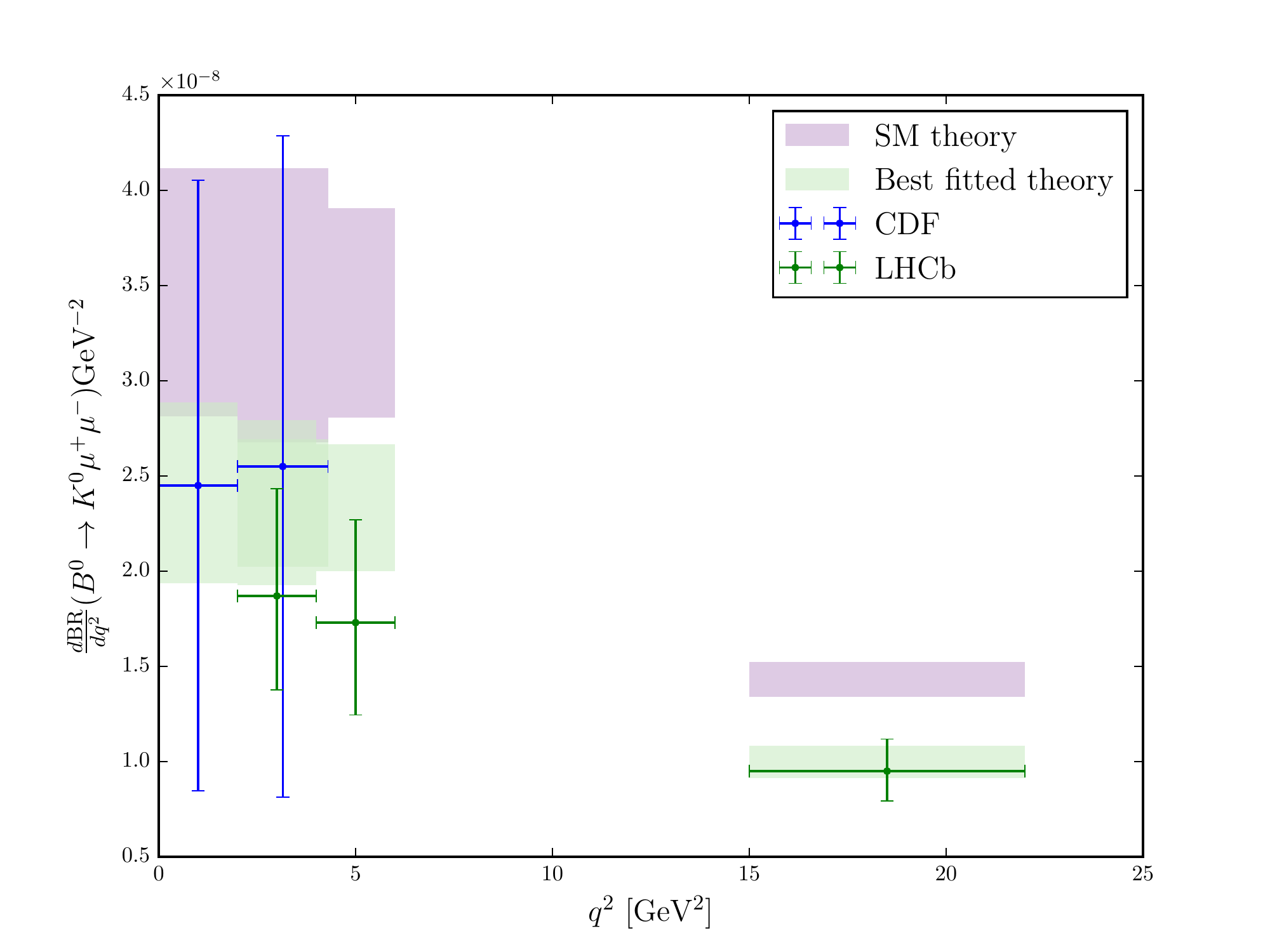}
		\includegraphics[width=0.4 \textwidth]{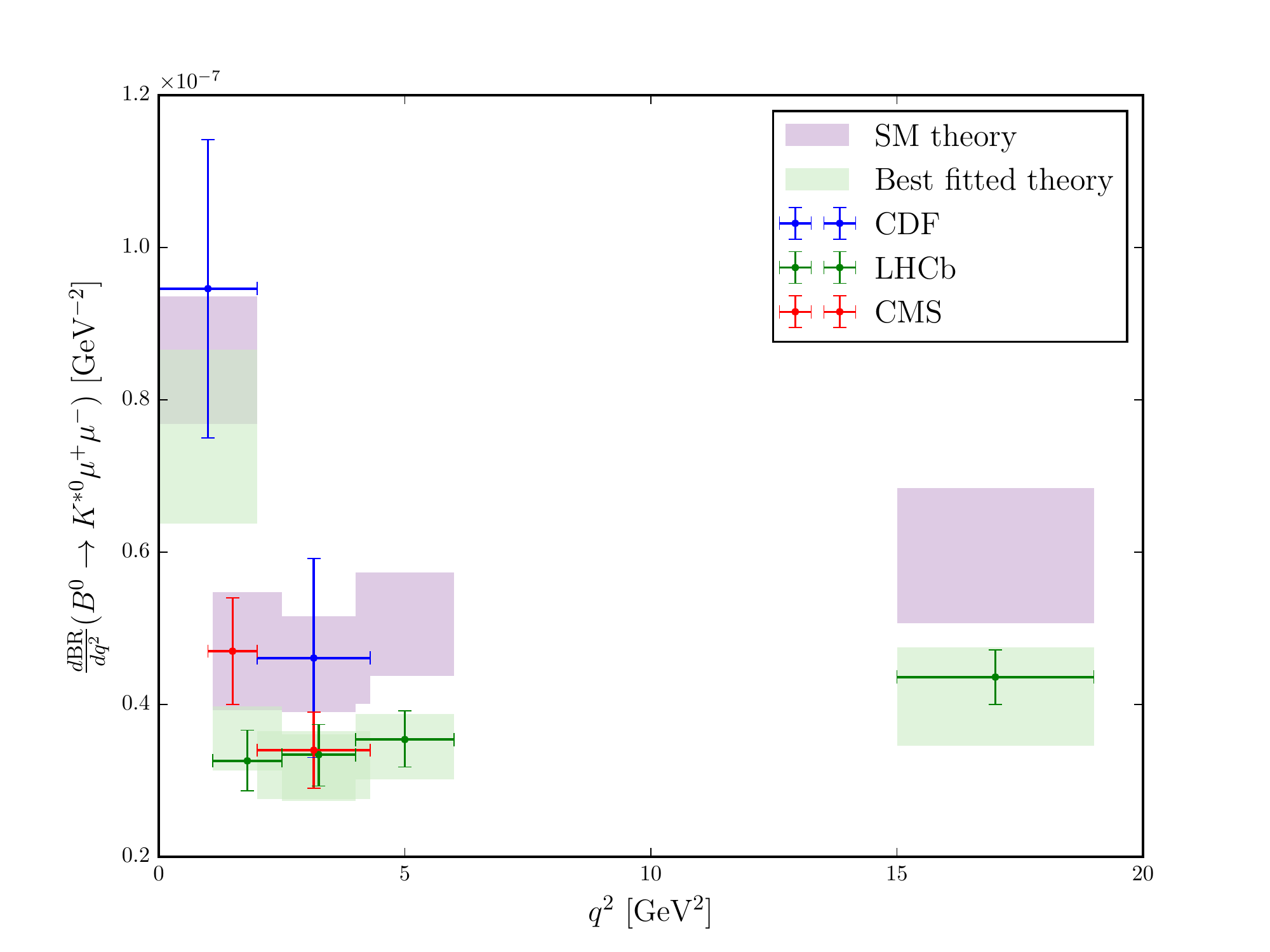}\\
		\includegraphics[width=0.4 \textwidth]{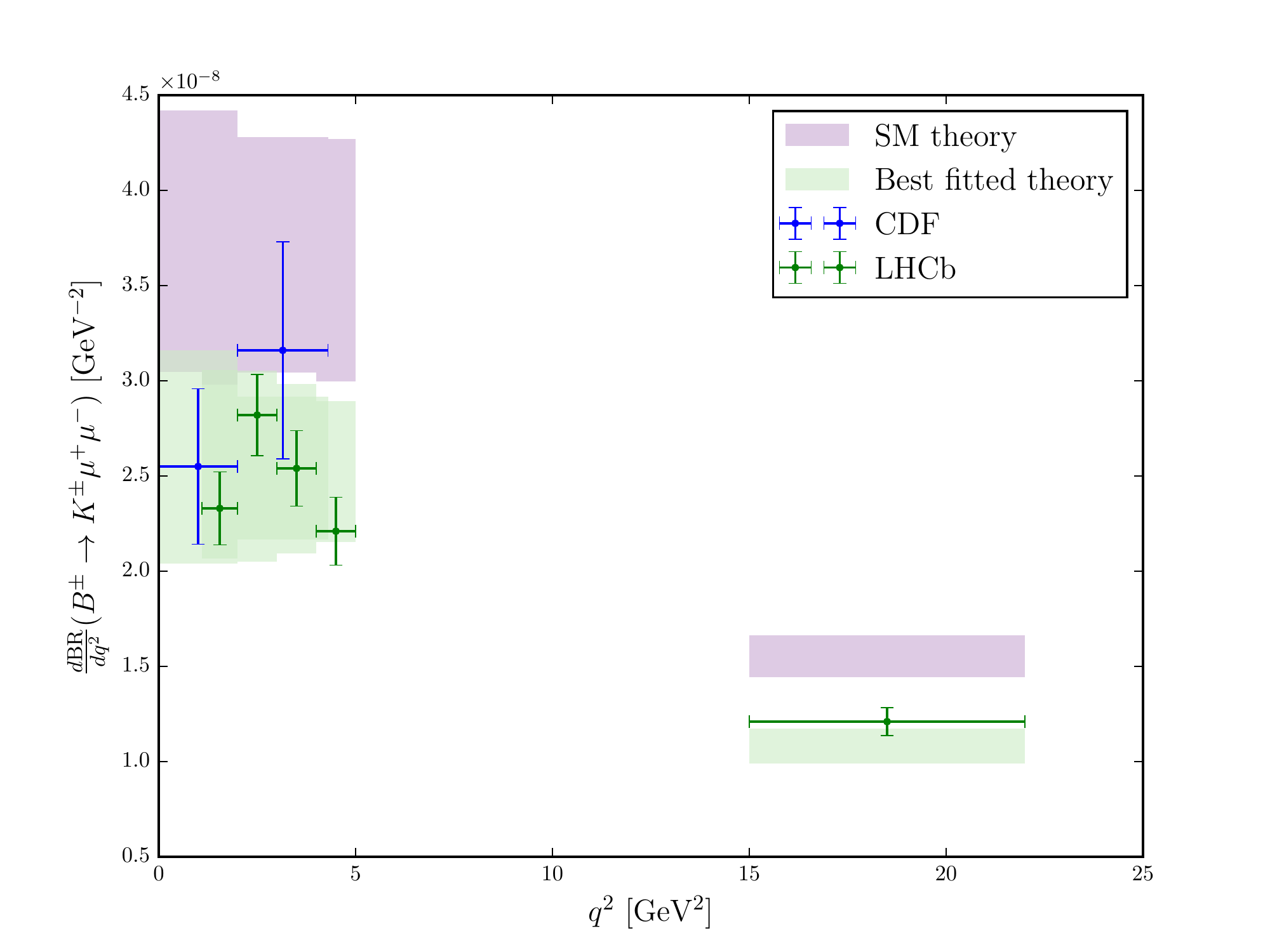}
		\includegraphics[width=0.4 \textwidth]{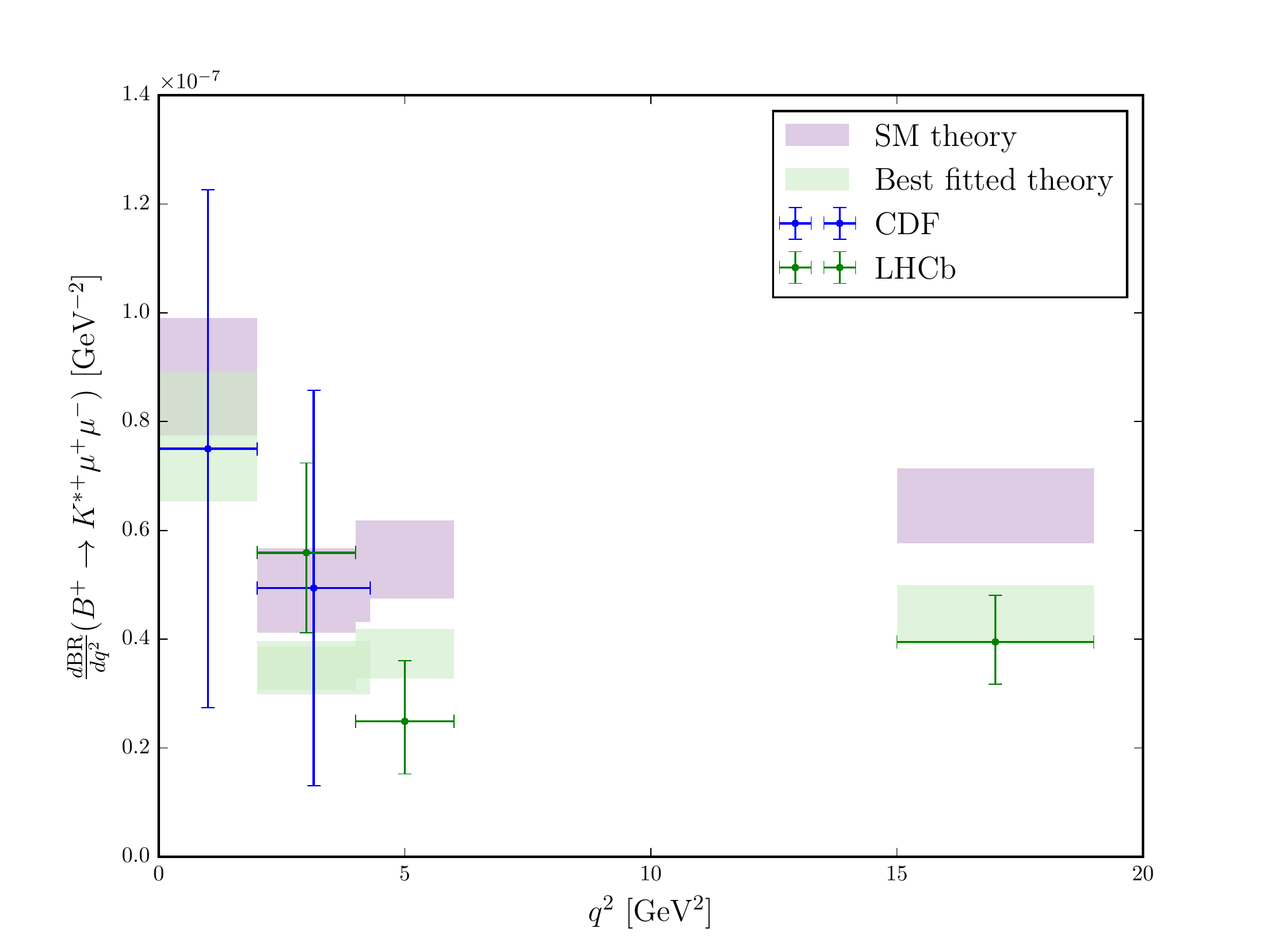}
	\end{center}
	\caption{The $B$ decay branching ratios ${\rm dBR}(B^0\to K^0 \mu^+ \mu^-)/{ dq^2}$ (upper left), ${\rm dBR}(B^0\to K^{*0} \mu^+ \mu^-)/{dq^2}$ (upper right), ${\rm dBR}(B^\pm\to K^\pm \mu^+ \mu^-)/{dq^2}$ (lower left), ${\rm dBR}(B^+\to K^{*+} \mu^+ \mu^-)/{dq^2}$  (lower right) for different energy bins. The experimental data are expressed by dots with uncertainties. The purple and green bands represent the theoretical predictions of the SM and our model with the best fitted $C_9$, respectively. The theoretical uncertainties are calculated by varying the input parameters with the Gaussian uncertainties.}  
	\label{fig:RK_data} 	
\end{figure}

We first consider the effective operators for the $R_{K}$ anomaly in terms of $X_q \equiv v_{\Phi}Y_{q}/\sqrt{2}m_{Q}~(q=b,s)$ and $X_\mu \equiv v_{\Phi}Y_{\mu}/\sqrt{2}m_{L}$. With these expressions, the Eq. (\ref{C9_new}) is described by
\begin{align}
C_{9, \rm new}(=-C_{10, \rm new})=-\frac{1}{2v^2_{\Phi}|C^{bs}_{\rm SM}|}X_bX_sX^2_{\mu}.
\end{align}
This expression consists of only relevant model parameters through the combinations of $X_q$ and $X_{\mu}$ that are related to the mixing angles in the quark and lepton sectors. It should be noted that the value of $C_{9, \rm new}$ in Table. \ref{table:model_analysis} is positive, therefore either $X_b$ or $X_s$ should be negative to compensate for the minus sign in the above expression. The left plot in Fig. \ref{fig:RK_anomaly} shows $1$$\sigma$ and $2$$\sigma$ regions with black solid and dashed lines in $(-X_bX_s,~X_{\mu}^2)$ plane with the fixed value of $v_{\Phi}=700~$GeV.
The best fitted value is drawn by a green line, and the blue region is excluded by $B_s-\bar{B_s}$ mixing. We also consider the constraint from the $B\to K\nu\nu$ process, however, the present parameter region can evade the restriction. It is seen that $X_bX_s$ is tightly constrained by the $B_s-\bar{B_s}$ mixing and the allowed size is roughly $O(10^{-3})$ at most. And, it follows that a somewhat large value of $X_{\mu}^2\sim O(0.1)$ is necessary to achieve the required Wilson coefficients for the $R_{K}$. 
Such a large value can be obtained by taking a large $Y_{\mu}$ or a small $m_L$. Here, we  fix the mass $m_L$ at 1.4 TeV and see how large Yukawa coupling, $Y_{\mu}$, is needed at one benchmark region of $-X_bX_s=2.3\times 10^{-3}$. The region is indicated by a red line in the left panel in Fig. \ref{fig:RK_anomaly}, and the benchmark value results in $Y_bY_s=-0.0184$ if $m_Q=m_L$. Since our assumption is $|Y_b|\gg |Y_s|$, we simply take $Y_b=-1$ and $Y_s=0.0184$.  In the right plot of Fig. \ref{fig:RK_anomaly}, the size of $C_{9,\rm new}$ is shown as a function of $Y_{\mu}$. Here, instead of the mass-insertion method, we use $g^L_{sb}$ and $g^L_{\mu\mu}$ obtained after diagonalization of the mass matrices. All lines in the right figure correspond to those in the left one. The $2\sigma$ region requires $1.2\leq Y_{\mu}\leq 1.9$, and the best fitted value stays around $Y_{\mu}\sim 1.5$. The values of $Y_b=-1,~Y_s=0.0184$ and $Y_{\mu}=1.5$ yield $g^L_{sb}/g^{\prime}=-2.17\times 10^{-3}$ and $g^L_{\mu\mu}/g^{\prime}=0.22$, which leads to $C_{9,\rm new}=0.6$. The current size of $g^L_{\mu\mu}/m_{Z^{\prime}}$ gives the cross section of the neutrino trident production in Eq. (\ref{NT}), $\sigma_{\rm SM+NP}/\sigma_{\rm SM}=1.0$, which is within the $1\sigma$ error.  Later, we also use these Yukawa couplings as the benchmark values for the discussion of the $R_D$ anomaly.

Figure~\ref{fig:RK_data} shows the experimental data and theoretical predictions on the $B$ decay branching ratios in different energy bins for  ${\text{dBR}}(B^0\to K^0 \mu^+ \mu^-)/{dq^2}$ (upper left), ${\rm dBR}(B^0\to K^{*0} \mu^+ \mu^-)/{dq^2}$ (upper right), ${\rm dBR}(B^\pm\to K^\pm \mu^+ \mu^-)/{dq^2}$ (lower left), ${\rm dBR}(B^+\to K^{*+} \mu^+ \mu^-)/{dq^2}$  (lower right). The experimental results obtained by CDF~\cite{CDF:2012qwd}, LHCb~\cite{Aaij:2016flj}, CMS~\cite{Chatrchyan:2013cda,Khachatryan:2015isa} are plotted.
The purple bands correspond to the SM predictions with uncertainties obtained by varying the input parameters with the Gaussian distribution. The green bands represent the best fitted value of $C_{9,\rm new}$, which is indicated by the green line in Fig. \ref{fig:RK_anomaly}, with the uncertainties obtained by the same way as the purple bands. The width of the purple and green bands are not the same, and their ratios are roughly proportional to $|C^{bs}_{\rm SM}/(C^{bs}_{\rm SM}+C_i)|^2~(i=9,10)$.  It is seen that our model provides a better fit on the data than the SM.

\begin{figure}[t]
	\begin{center}
		\includegraphics[width=0.4 \textwidth]{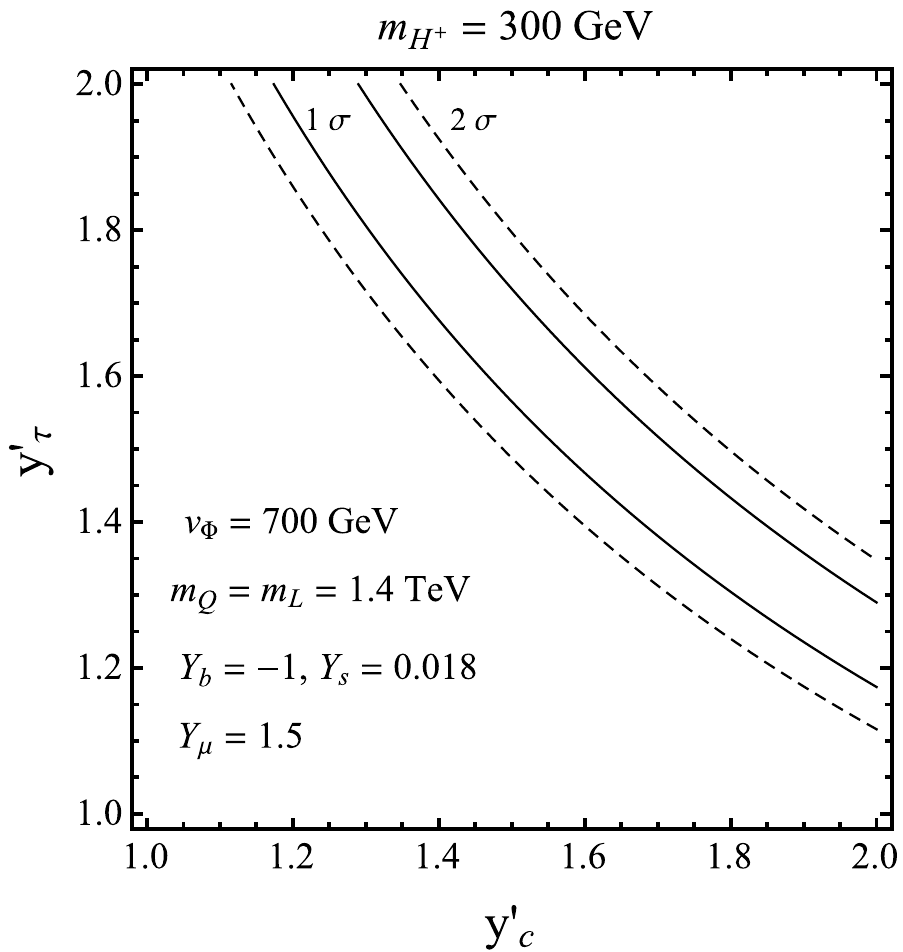}
	\end{center}
	\caption{$1\sigma$ and $2\sigma$ regions of $C_S^{\prime}$ with black solid and dashed lines, respectively. It is taken that $m_{H'}=300~$GeV, and the benchmark parameters
	for the $R_K$ anomaly: $v_{\Phi}=700~$GeV,~$m_Q=m_L=1.4~$TeV, $Y_b=-1,~Y_s=0.018$ and $Y_{\mu}=1.5$.}
	\label{fig:RD_anomaly} 
\end{figure}

Finally, we discuss the possibility of explaining the $R_{D}$ anomaly using the parameter space which could explain the $R_K$ anomaly. 
Once $Y_b$ and $Y_s$ are fixed, $Y_c$ is also given due to the $SU(2)$ symmetry. Using the benchmark values, $Y_b=-1,~Y_s=0.018$, we find that  $Y_c=-0.02$ where the $Y_b$ part dominantly gives a contribution to $Y_c$ since $(V_{\rm CKM})_{cb} Y_b>Y_s$. As seen in Fig. \ref{fig:cscsp}, positive $C_S^{\prime}$ is disfavored by the upper limit on the $B^-_c\to \tau^- \bar{\nu}$ process, and the absolute values of $C_S$ and $C_S^{\prime}$ should be $O(0.1)$. In the current setup, the negative $C_S^{\prime}$ can be realized by $Y_b=-1$. However, a naive estimation implies that $C_S^{\prime} \gg C_S$ since $C_S$ is proportional to $Y_c$ which is two orders of magnitude smaller than $Y_b$. Therefore, we exclusively focus on a specific situation with $C_S^{\prime}\neq 0$ but $C_S=0$ as listed in Table \ref{table:model_analysis}.

Figure \ref{fig:RD_anomaly} shows the $1\sigma$ and $2\sigma$ regions of $C_S^{\prime}$ in $(y^{\prime}_c,y^{\prime}_{\tau})$ plane. Some of the model parameters 
 are correlated with the $R_K$ anomaly, and they are fixed at $Y_b=-1,~Y_s=0.018,~Y_{\mu}=1.5,~v_{\phi}=700~$GeV and $m_Q=m_L=1.4~$TeV. The charged scalar mass is taken as $300~$GeV. As we can expect from the required value of $C_S^{\prime}$, it is found that the large Yukawa couplings $y_c^{\prime}$ and $y_{\tau}^{\prime}$ are needed for the explanation of the $R_D$ anomaly. As long as we focus on only $C_S^{\prime}$, the size of $y_b^{\prime}$ is not predicted. Therefore, our current setup does not affect $b\to s\gamma$ process. However, it is found that, 
even if we take $y_c^{\prime}=2$ and $y_b^{\prime}=1$,  $C_7(\mu_b)$ roughly becomes $1.6\times 10^{-3}$ which is consistent with the current observed value.


\section{Conclusion}
\label{sec:conclude}

We have studied a hidden gauged $U(1)_H$ extension of the SM including vector-like fermion (and scalar) mediators and investigated possibilities of explanations for the existing $R_K$ (and $R_D$) anomalies in $B$ physics. 
The vector-like fermions have the same SM charges as the SM doublet fermions, and  they mix with the left-handed SM fermions. 
The new scalar is an inert doublet, which possesses the Yukawa interaction between the left-handed vector-like fermions and the right-handed SM fermions. 
While the vector-like fermions play a role in inducing $b\to s \mu\mu$ transition through the $Z^{\prime}$ exchange, the inert doublet yields significant charged scalar couplings for $b\to c\tau\nu$ transition. 

In order to determine desired sizes of the Wilson coefficients, we utilized the flavio package and performed a global Bayesian fit including several observables, $B^{0, \pm}\to K^{*0,\pm}\mu\mu, B_s\to\phi\mu\mu, B_s\to X_s\mu\mu $ and $ B_s\to \mu\mu$.  At the same time, we also considered various flavor constraints, $B_s-\bar{B_s}$ mixing, $B \to K \nu\bar{\nu}$, neutrino trident productions, $b \to s \gamma$, and $B^-_c\to\tau^-\bar{\nu}$. It is found that the absolute values of $C_9 (= - C_{10})$, $C_{S}^{\prime}$ and $C_{S}$ are required to be $O(0.5)$ roughly. Also, it turned out that the positive values of $C_{S}^{\prime}$ are almost excluded by  $B^-_c\to\tau^- \bar{\nu}$ if we impose its severe limit on the branching ratio, ${\rm BR}(B^-_c\to\tau^-\bar{\nu})<10\%$.

The required sizes of $C_9$ and $C_{10}$ can be achieved by somewhat large $Y_\mu$, which describes the mixing parameter between the vector-like and SM leptons, since those in the quark sector $Y_b$ and $Y_s$ are highly constrained by the $B_s-\bar{B}_s$ mixing. Our benchmark point with $m_Q=m_L=1.4~$TeV, $v_{\Phi}=700~$GeV and $Y_bY_s=-0.018$ needs $Y_{\mu} \sim 1.5$ for the successful scenario of the $R_K$ anomaly. This benchmark leads to a consistent situation in which $C_S$ is suppressed due to tiny $Y_c$ and negative $C_S^{\prime}$ is obtained by taking $Y_b<0$. The possible region of $C_S^{\prime}$ requires $O(1)$ couplings of $y_c^{\prime}$ and $y_{\tau}^{\prime}$.

Since this model contains new particles, there are promising collider and astrophysical signatures. Current LHC searches on $pp \to \mu\bar{\mu}$ and $pp \to 4 \mu$ put strong constraints on the $Z'$ masses, although the light region of $M_{Z'} < 100$ GeV is still allowed. Furthermore, since the $U(1)$ gauge symmetry is hidden, it is likely that other hidden particles also exist, such as dark matter candidate, in the hidden sector. 
Although the hidden sector particles do not contribute to the $B$ physics observables, 
they are able to contribute to the dark matter relic density and possible indirect detections, 
and would relax the constraints on the heavy $Z'$.  On the other hand,
the charged Higgs is not tightly constrained due to its Yukawa texture. 
We anticipate that the future collider searches will explore larger mass regions for the $Z'$ and charged Higgs, and these collider signatures and 
potential dark matter signatures could be able to validate this model.




\renewcommand{\arraystretch}{0.7}
\begin{table}[h]
\begin{center}
\small\begin{tabular}{||c|c|c||}

\hline\hline
  \multicolumn{3}{|c|}{\textbf{Branching ratios}}   \\
\hline\hline

\cline{1-3}
{{Observable}}  & {{$[q^{2}_{\text{min}},q^{2}_{\text{max}}]$ [GeV$^{2}$]}} & {{Experiments}} \\ \hline

$\frac{d}{dq^{2}}\text{BR}(B\to X_{s} \mu\mu)$ & [1, 6],[14.2, 25] & BaBar~\cite{Lees:2013nxa} \\ 
\hline
\multirow{2}{*}{$\frac{d}{dq^{2}}\text{BR}(B^{+}\to K^{*+}\mu\mu)$} & [0.0, 2.0], [2.0, 4.3]&CDF~\cite{CDF:2012qwd}\\ \cline{2-3}
&[2,4], [4,6], [15,19] &LHCb~\cite{Aaij:2014pli}\\  \hline

\multirow{2}{*}{$\frac{d}{dq^{2}}\text{BR}(B^{\pm}\to K^{\pm}\mu\mu)$} & [0.0, 2.0], [2.0, 4.3]&CDF~\cite{CDF:2012qwd}\\ \cline{2-3}
&[1.1, 2], [2, 3], [3, 4], [4, 5], [15,22] &LHCb~\cite{Aaij:2014pli}\\  \hline

\multirow{3}{*}{$\frac{d}{dq^{2}}\text{BR}(B^{0}\to K^{*0}\mu\mu)$} & [0.0, 2.0], [2.0, 4.3]&CDF~\cite{CDF:2012qwd}\\ \cline{2-3}
&[1.1, 2.5], [2.5, 4], [4, 6], [15, 19] &LHCb~\cite{Aaij:2016flj}\\  
\cline{2-3}
&[1, 2], [2, 4.3] & CMS~\cite{Chatrchyan:2013cda,Khachatryan:2015isa}\\
\hline

\multirow{2}{*}{$\frac{d}{dq^{2}}\text{BR}(B^{0}\to K^{0}\mu\mu)$} & [0.0, 2.0], [2.0, 4.3]&CDF~\cite{CDF:2012qwd}\\ \cline{2-3}
&[2.5, 4], [4, 6], [15,22] &LHCb~\cite{Aaij:2014pli}\\  \hline

\multirow{2}{*}{$\frac{d}{dq^{2}}\text{BR}(B_s\to \phi\mu\mu)$} & [1, 6] &CDF~\cite{CDF:2012qwd}\\ \cline{2-3}
&[1, 6], [15, 19]&LHCb~\cite{Aaij:2015esa}\\  \hline

$R_{K^*}$ & [0.045, 1.1], [1.1, 6.0] & LHCb~\cite{Aaij:2017vbb}\\
\hline
$R_{K}$ & [1.0, 6.0] & LHCb~\cite{Aaij:2014ora}\\
\hline

$BR(B_s\to\mu\mu),~ BR(B^0\to \mu\mu)$ & -----& LHCb\cite{Tolk:2017dqr}\\
\hline

  \multicolumn{3}{|c|}{\textbf{Angular Observables}}   \\
\hline\hline

\multirow{3}{*}{$\langle A_{FB} \rangle (B^{0}\to K^{*0}\mu\mu)$} & [0.0, 2.0], [2.0, 4.3]& CDF~\cite{CDF:2012qwd}\\ \cline{2-3}
&[1.1, 2.5], [2.5, 4], [4, 6], [15, 19]&LHCb~\cite{Aaij:2015oid}\\  
\cline{2-3}
&[1, 2], [2, 4.3], [4.3, 6] & CMS~\cite{CMS:2017ivg,Chatrchyan:2013cda,Khachatryan:2015isa}\\
\hline

\multirow{4}{*}{$\langle F_{L} \rangle (B^{0}\to K^{*0}\mu\mu)$} & [0.0, 2.0], [2.0, 4.3]& CDF~\cite{CDF:2012qwd}\\ \cline{2-3}
&[1.1, 2.5], [2.5, 4], [4, 6], [15, 19]&LHCb~\cite{Aaij:2015oid}\\  
\cline{2-3}
&[1, 2], [2, 4.3], [4.3, 6] & CMS~\cite{CMS:2017ivg,Chatrchyan:2013cda,Khachatryan:2015isa}\\
\cline{2-3}
& [0.04, 2.0], [2.0, 4.0], [4.0, 6.0]& ATLAS~\cite{ATLAS:2017dlm}\\
\hline

\multirow{2}{*}{$\langle S_{3} \rangle (B^{0}\to K^{*0}\mu\mu)$} 
&[1.1, 2.5], [2.5, 4], [4, 6], [15, 19]&LHCb~\cite{Aaij:2015oid}\\  
\cline{2-3}
&[1, 2], [2, 4.3], [4.3, 6] & CMS~\cite{CMS:2017ivg,Chatrchyan:2013cda,Khachatryan:2015isa}\\
\hline

\multirow{2}{*}{$\langle S_{4} \rangle (B^{0}\to K^{*0}\mu\mu)$} 
&[1.1, 2.5], [2.5, 4], [4, 6], [15, 19]&LHCb~\cite{Aaij:2015oid}\\  
\cline{2-3}
&[1, 2], [2, 4.3], [4.3, 6] & CMS~\cite{CMS:2017ivg,Chatrchyan:2013cda,Khachatryan:2015isa}\\
\hline

\multirow{2}{*}{$\langle S_{5} \rangle (B^{0}\to K^{*0}\mu\mu)$} 
&[1.1, 2.5], [2.5, 4], [4, 6], [15, 19]&LHCb~\cite{Aaij:2015oid}\\  
\cline{2-3}
&[1, 2], [2, 4.3], [4.3, 6] & CMS~\cite{CMS:2017ivg,Chatrchyan:2013cda,Khachatryan:2015isa}\\
\hline

\multirow{3}{*}{$\langle P_{1} \rangle (B^{0}\to K^{*0}\mu\mu)$} & [0.04, 2.0], [2.0, 4.0], [4.0, 6.0]& ATLAS~\cite{ATLAS:2017dlm}\\ \cline{2-3}
&[1.1, 2.5], [2.5, 4], [4, 6], [15, 19]&LHCb~\cite{Aaij:2015oid}\\  
\cline{2-3}
&[1, 2], [2, 4.3], [4.3, 6] & CMS~\cite{CMS:2017ivg,Chatrchyan:2013cda,Khachatryan:2015isa}\\
\hline

\multirow{2}{*}{$\langle P'_{4} \rangle (B^{0}\to K^{*0}\mu\mu)$} & [0.04, 2.0], [2.0, 4.0], [4.0, 6.0]& ATLAS~\cite{ATLAS:2017dlm}\\ \cline{2-3}
&[1.1, 2.5], [2.5, 4], [4, 6], [15, 19]&LHCb~\cite{Aaij:2015oid}\\  
\cline{2-3}
\hline

\multirow{3}{*}{$\langle P'_{5} \rangle (B^{0}\to K^{*0}\mu\mu)$} & [0.04, 2.0], [2.0, 4.0], [4.0, 6.0]& ATLAS~\cite{ATLAS:2017dlm}\\ \cline{2-3}
&[1.1, 2.5], [2.5, 4], [4, 6], [15, 19]&LHCb~\cite{Aaij:2015oid}\\  
\cline{2-3}
&[1, 2], [2, 4.3], [4.3, 6] & CMS~\cite{CMS:2017ivg,Chatrchyan:2013cda,Khachatryan:2015isa}\\
\hline

$\langle \overline{F_L} \rangle (B^{0}\to K^{*0}\mu\mu)$ & &\\
$\langle \overline{S_3} \rangle (B^{0}\to K^{*0}\mu\mu)$& [2, 5], [15, 19] &LHCb~\cite{Aaij:2015oid} \\
 $\langle \overline{S_4} \rangle (B^{0}\to K^{*0}\mu\mu)$ & &\\
\hline

   \end{tabular}
\end{center}
\caption{List of observable used in global fit of $R_{K^{(*)}}$ anomaly.
\label{tab:rk}}
\end{table}%
\normalsize

\begin{acknowledgments}
HL thanks David Straub for helpful discussions about flavio and Aniket Jorglekar for valuable discussion on the diagonalization of fermion mass matrix. JHY and KF are supported by DOE Grant DE-SC0011095. 
\end{acknowledgments}




\begin{thebibliography}{99}

\bibitem{Aaij:2017vbb} 
  R.~Aaij {\it et al.} [LHCb Collaboration],
  JHEP {\bf 1708}, 055 (2017)
  doi:10.1007/JHEP08(2017)055
  [arXiv:1705.05802 [hep-ex]].

\bibitem{Bordone:2016gaq} 
  M.~Bordone, G.~Isidori and A.~Pattori,
  Eur.\ Phys.\ J.\ C {\bf 76}, no. 8, 440 (2016)
  doi:10.1140/epjc/s10052-016-4274-7
  [arXiv:1605.07633 [hep-ph]].

\bibitem{Descotes-Genon:2015uva} 
  S.~Descotes-Genon, L.~Hofer, J.~Matias and J.~Virto,
  JHEP {\bf 1606}, 092 (2016)
  doi:10.1007/JHEP06(2016)092
  [arXiv:1510.04239 [hep-ph]].

\bibitem{Capdevila:2016ivx} 
  B.~Capdevila, S.~Descotes-Genon, J.~Matias and J.~Virto,
  JHEP {\bf 1610}, 075 (2016)
  doi:10.1007/JHEP10(2016)075
  [arXiv:1605.03156 [hep-ph]].

\bibitem{Capdevila:2017ert} 
  B.~Capdevila, S.~Descotes-Genon, L.~Hofer and J.~Matias,
  JHEP {\bf 1704}, 016 (2017)
  doi:10.1007/JHEP04(2017)016
  [arXiv:1701.08672 [hep-ph]].
  
\bibitem{Serra:2016ivr} 
  N.~Serra, R.~Silva Coutinho and D.~van Dyk,
  Phys.\ Rev.\ D {\bf 95}, no. 3, 035029 (2017)
  doi:10.1103/PhysRevD.95.035029
  [arXiv:1610.08761 [hep-ph]].

\bibitem{Dvan}  
D. van Dyk et al.. EOS — A HEP program for flavor observables. \url{https://eos.github.io}  

\bibitem{Straub:2015ica} 
  A.~Bharucha, D.~M.~Straub and R.~Zwicky,
  JHEP {\bf 1608}, 098 (2016)
  doi:10.1007/JHEP08(2016)098
  [arXiv:1503.05534 [hep-ph]].

\bibitem{Altmannshofer:2017fio} 
  W.~Altmannshofer, C.~Niehoff, P.~Stangl and D.~M.~Straub,
  Eur.\ Phys.\ J.\ C {\bf 77}, no. 6, 377 (2017)
  doi:10.1140/epjc/s10052-017-4952-0
  [arXiv:1703.09189 [hep-ph]].

\bibitem{flavio}  
D. Straub, A Python package for flavour physics phenomenology in the Standard Model and beyond. \url{https://flav-io.github.io}

\bibitem{Jager:2014rwa} 
  S.~Jager and J.~Martin Camalich,
  Phys.\ Rev.\ D {\bf 93}, no. 1, 014028 (2016)
  doi:10.1103/PhysRevD.93.014028
  [arXiv:1412.3183 [hep-ph]].


\bibitem{Aaij:2014ora} 
  R.~Aaij {\it et al.} [LHCb Collaboration],
  Phys.\ Rev.\ Lett.\  {\bf 113}, 151601 (2014)
  doi:10.1103/PhysRevLett.113.151601
  [arXiv:1406.6482 [hep-ex]].
  
\bibitem{Lees:2012xj} 
  J.~P.~Lees {\it et al.} [BaBar Collaboration],
  Phys.\ Rev.\ Lett.\  {\bf 109}, 101802 (2012)
  doi:10.1103/PhysRevLett.109.101802
  [arXiv:1205.5442 [hep-ex]].

\bibitem{Lees:2013uzd} 
  J.~P.~Lees {\it et al.} [BaBar Collaboration],
  Phys.\ Rev.\ D {\bf 88}, no. 7, 072012 (2013)
  doi:10.1103/PhysRevD.88.072012
  [arXiv:1303.0571 [hep-ex]].
  
\bibitem{Huschle:2015rga} 
  M.~Huschle {\it et al.} [Belle Collaboration],
  Phys.\ Rev.\ D {\bf 92}, no. 7, 072014 (2015)
  doi:10.1103/PhysRevD.92.072014
  [arXiv:1507.03233 [hep-ex]].

\bibitem{Sato:2016svk} 
  Y.~Sato {\it et al.} [Belle Collaboration],
  Phys.\ Rev.\ D {\bf 94}, no. 7, 072007 (2016)
  doi:10.1103/PhysRevD.94.072007
  [arXiv:1607.07923 [hep-ex]].

\bibitem{Hirose:2016wfn} 
  S.~Hirose {\it et al.} [Belle Collaboration],
  Phys.\ Rev.\ Lett.\  {\bf 118}, no. 21, 211801 (2017)
  doi:10.1103/PhysRevLett.118.211801
  [arXiv:1612.00529 [hep-ex]].

\bibitem{Aaij:2015yra} 
  R.~Aaij {\it et al.} [LHCb Collaboration],
  Phys.\ Rev.\ Lett.\  {\bf 115}, no. 11, 111803 (2015)
  Erratum: [Phys.\ Rev.\ Lett.\  {\bf 115}, no. 15, 159901 (2015)]
  doi:10.1103/PhysRevLett.115.159901, 10.1103/PhysRevLett.115.111803
  [arXiv:1506.08614 [hep-ex]].


\bibitem{Fajfer:2012vx} 
  S.~Fajfer, J.~F.~Kamenik and I.~Nisandzic,
  Phys.\ Rev.\ D {\bf 85}, 094025 (2012)
  doi:10.1103/PhysRevD.85.094025
  [arXiv:1203.2654 [hep-ph]].
  
\bibitem{Bigi:2017jbd} 
  D.~Bigi, P.~Gambino and S.~Schacht,
  arXiv:1707.09509 [hep-ph].
  
\bibitem{Jaiswal:2017rve} 
  S.~Jaiswal, S.~Nandi and S.~K.~Patra,
  arXiv:1707.09977 [hep-ph].
  
\bibitem{Lattice:2015rga} 
  J.~A.~Bailey {\it et al.} [MILC Collaboration],
  Phys.\ Rev.\ D {\bf 92}, no. 3, 034506 (2015)
  doi:10.1103/PhysRevD.92.034506
  [arXiv:1503.07237 [hep-lat]].
  
\bibitem{Na:2015kha} 
  H.~Na {\it et al.} [HPQCD Collaboration],
  Phys.\ Rev.\ D {\bf 92}, no. 5, 054510 (2015)
  Erratum: [Phys.\ Rev.\ D {\bf 93}, no. 11, 119906 (2016)]
  doi:10.1103/PhysRevD.93.119906, 10.1103/PhysRevD.92.054510
  [arXiv:1505.03925 [hep-lat]].

\bibitem{Bigi:2016mdz} 
  D.~Bigi and P.~Gambino,
  Phys.\ Rev.\ D {\bf 94}, no. 9, 094008 (2016)
  doi:10.1103/PhysRevD.94.094008
  [arXiv:1606.08030 [hep-ph]].

%
\bibitem{Altmannshofer:2016jzy} 
  W.~Altmannshofer, S.~Gori, S.~Profumo and F.~S.~Queiroz,
  JHEP {\bf 1612}, 106 (2016)
  doi:10.1007/JHEP12(2016)106
  [arXiv:1609.04026 [hep-ph]].

\bibitem{Ko:2017lzd}
  P.~Ko, Y.~Omura, Y.~Shigekami and C.~Yu,
  Phys.\ Rev.\ D {\bf 95} (2017) no.11,  115040
  doi:10.1103/PhysRevD.95.115040
  [arXiv:1702.08666 [hep-ph]].
  
\bibitem{Datta:2017pfz} 
  A.~Datta, J.~Liao and D.~Marfatia,
  Phys.\ Lett.\ B {\bf 768}, 265 (2017)
  doi:10.1016/j.physletb.2017.02.058
  [arXiv:1702.01099 [hep-ph]].

  
%
\bibitem{Datta:2017ezo} 
  A.~Datta, J.~Kumar, J.~Liao and D.~Marfatia,
  arXiv:1705.08423 [hep-ph].

\bibitem{Bhattacharya:2016mcc} 
  B.~Bhattacharya, A.~Datta, J.~P.~Guevin, D.~London and R.~Watanabe,
  JHEP {\bf 1701}, 015 (2017)
  doi:10.1007/JHEP01(2017)015
  [arXiv:1609.09078 [hep-ph]].

\bibitem{Cline:2017ihf} 
  J.~M.~Cline and J.~Martin Camalich,
  Phys.\ Rev.\ D {\bf 96}, no. 5, 055036 (2017)
  doi:10.1103/PhysRevD.96.055036
  [arXiv:1706.08510 [hep-ph]].
  
\bibitem{Faisel:2017glo} 
  G.~Faisel and J.~Tandean,
  arXiv:1710.11102 [hep-ph].

\bibitem{Hiller:2014yaa} 
  G.~Hiller and M.~Schmaltz,
  Phys.\ Rev.\ D {\bf 90}, 054014 (2014)
  doi:10.1103/PhysRevD.90.054014
  [arXiv:1408.1627 [hep-ph]].
  
\bibitem{Duraisamy:2016gsd} 
  M.~Duraisamy, S.~Sahoo and R.~Mohanta,
  Phys.\ Rev.\ D {\bf 95}, no. 3, 035022 (2017)
  doi:10.1103/PhysRevD.95.035022
  [arXiv:1610.00902 [hep-ph]].


\bibitem{Chen:2017hir} 
  C.~H.~Chen, T.~Nomura and H.~Okada,
  Phys.\ Lett.\ B {\bf 774}, 456 (2017)
  doi:10.1016/j.physletb.2017.10.005
  [arXiv:1703.03251 [hep-ph]].
  
\bibitem{Aloni:2017ixa} 
  D.~Aloni, A.~Dery, C.~Frugiuele and Y.~Nir,
  JHEP {\bf 1711}, 109 (2017)
  doi:10.1007/JHEP11(2017)109
  [arXiv:1708.06161 [hep-ph]].

\bibitem{Calibbi:2017qbu} 
  L.~Calibbi, A.~Crivellin and T.~Li,
  arXiv:1709.00692 [hep-ph].


%
\bibitem{Bailey:2012jg} 
  J.~A.~Bailey {\it et al.},
  Phys.\ Rev.\ Lett.\  {\bf 109}, 071802 (2012)
  doi:10.1103/PhysRevLett.109.071802
  [arXiv:1206.4992 [hep-ph]].

\bibitem{Celis:2012dk} 
  A.~Celis, M.~Jung, X.~Q.~Li and A.~Pich,
  JHEP {\bf 1301}, 054 (2013)
  doi:10.1007/JHEP01(2013)054
  [arXiv:1210.8443 [hep-ph]].

\bibitem{Tanaka:2012nw} 
  M.~Tanaka and R.~Watanabe,
  Phys.\ Rev.\ D {\bf 87}, no. 3, 034028 (2013)
  doi:10.1103/PhysRevD.87.034028
  [arXiv:1212.1878 [hep-ph]].

\bibitem{Crivellin:2012ye} 
  A.~Crivellin, C.~Greub and A.~Kokulu,
  Phys.\ Rev.\ D {\bf 86}, 054014 (2012)
  doi:10.1103/PhysRevD.86.054014
  [arXiv:1206.2634 [hep-ph]].
  
\bibitem{Crivellin:2015hha} 
  A.~Crivellin, J.~Heeck and P.~Stoffer,
  Phys.\ Rev.\ Lett.\  {\bf 116}, no. 8, 081801 (2016)
  doi:10.1103/PhysRevLett.116.081801
  [arXiv:1507.07567 [hep-ph]].
  
\bibitem{Cline:2015lqp} 
  J.~M.~Cline,
  Phys.\ Rev.\ D {\bf 93}, no. 7, 075017 (2016)
  doi:10.1103/PhysRevD.93.075017
  [arXiv:1512.02210 [hep-ph]].
  
\bibitem{Alonso:2016oyd} 
  R.~Alonso, B.~Grinstein and J.~Martin Camalich,
  Phys.\ Rev.\ Lett.\  {\bf 118}, no. 8, 081802 (2017)
  doi:10.1103/PhysRevLett.118.081802
  [arXiv:1611.06676 [hep-ph]].

\bibitem{Iguro:2017ysu} 
  S.~Iguro and K.~Tobe,
  arXiv:1708.06176 [hep-ph].


%
\bibitem{He:2012zp} 
  X.~G.~He and G.~Valencia,
  Phys.\ Rev.\ D {\bf 87}, no. 1, 014014 (2013)
  doi:10.1103/PhysRevD.87.014014
  [arXiv:1211.0348 [hep-ph]].

\bibitem{Boucenna:2016qad} 
  S.~M.~Boucenna, A.~Celis, J.~Fuentes-Martin, A.~Vicente and J.~Virto,
  JHEP {\bf 1612}, 059 (2016)
  doi:10.1007/JHEP12(2016)059
  [arXiv:1608.01349 [hep-ph]].


%
\bibitem{Fajfer:2012jt} 
  S.~Fajfer, J.~F.~Kamenik, I.~Nisandzic and J.~Zupan,
  Phys.\ Rev.\ Lett.\  {\bf 109}, 161801 (2012)
  doi:10.1103/PhysRevLett.109.161801
  [arXiv:1206.1872 [hep-ph]].
  
\bibitem{Sakaki:2013bfa} 
  Y.~Sakaki, M.~Tanaka, A.~Tayduganov and R.~Watanabe,
  Phys.\ Rev.\ D {\bf 88}, no. 9, 094012 (2013)
  doi:10.1103/PhysRevD.88.094012
  [arXiv:1309.0301 [hep-ph]].

\bibitem{Dorsner:2013tla} 
  I.~Dorsner, S.~Fajfer, N.~Kosnik and I.~Nisandzic,
  JHEP {\bf 1311}, 084 (2013)
  doi:10.1007/JHEP11(2013)084
  [arXiv:1306.6493 [hep-ph]].

\bibitem{Li:2016vvp} 
  X.~Q.~Li, Y.~D.~Yang and X.~Zhang,
  JHEP {\bf 1608}, 054 (2016)
  doi:10.1007/JHEP08(2016)054
  [arXiv:1605.09308 [hep-ph]].
  
\bibitem{Crivellin:2017zlb} 
  A.~Crivellin, D.~Muller and T.~Ota,
  JHEP {\bf 1709}, 040 (2017)
  doi:10.1007/JHEP09(2017)040
  [arXiv:1703.09226 [hep-ph]].
  

\bibitem{Altmannshofer:2014cfa} 
  W.~Altmannshofer, S.~Gori, M.~Pospelov and I.~Yavin,
  Phys.\ Rev.\ D {\bf 89}, 095033 (2014)
  doi:10.1103/PhysRevD.89.095033
  [arXiv:1403.1269 [hep-ph]].
  
\bibitem{Altmannshofer:2014pba} 
  W.~Altmannshofer, S.~Gori, M.~Pospelov and I.~Yavin,
  Phys.\ Rev.\ Lett.\  {\bf 113}, 091801 (2014)
  doi:10.1103/PhysRevLett.113.091801
  [arXiv:1406.2332 [hep-ph]].
   
    


\bibitem{Yu:2016lof} 
  J.~H.~Yu,
  Phys.\ Rev.\ D {\bf 93}, no. 11, 113007 (2016)
  doi:10.1103/PhysRevD.93.113007
  [arXiv:1601.02609 [hep-ph]].

\bibitem{Ma:2006km} 
  E.~Ma,
  Phys.\ Rev.\ D {\bf 73}, 077301 (2006)
  doi:10.1103/PhysRevD.73.077301
  [hep-ph/0601225].
  
\bibitem{Fuyuto:2015gmk} 
  K.~Fuyuto, W.~S.~Hou and M.~Kohda,
  Phys.\ Rev.\ D {\bf 93}, no. 5, 054021 (2016)
  doi:10.1103/PhysRevD.93.054021
  [arXiv:1512.09026 [hep-ph]].

\bibitem{Poh:2017tfo} 
  Z.~Poh and S.~Raby,
  Phys.\ Rev.\ D {\bf 96}, no. 1, 015032 (2017)
  doi:10.1103/PhysRevD.96.015032
  [arXiv:1705.07007 [hep-ph]].
  

\bibitem{Altmannshofer:2013foa} 
  W.~Altmannshofer and D.~M.~Straub,
  Eur.\ Phys.\ J.\ C {\bf 73}, 2646 (2013)
  doi:10.1140/epjc/s10052-013-2646-9
  [arXiv:1308.1501 [hep-ph]].

\bibitem{Descotes-Genon:2013wba} 
  S.~Descotes-Genon, J.~Matias and J.~Virto,
  Phys.\ Rev.\ D {\bf 88}, 074002 (2013)
  doi:10.1103/PhysRevD.88.074002
  [arXiv:1307.5683 [hep-ph]].
  
\bibitem{Beaujean:2013soa} 
  F.~Beaujean, C.~Bobeth and D.~van Dyk,
  Eur.\ Phys.\ J.\ C {\bf 74}, 2897 (2014)
  Erratum: [Eur.\ Phys.\ J.\ C {\bf 74}, 3179 (2014)]
  doi:10.1140/epjc/s10052-014-2897-0, 10.1140/epjc/s10052-014-3179-6
  [arXiv:1310.2478 [hep-ph]].


\bibitem{Buchner:2014nha} 
  J.~Buchner {\it et al.},
  Astron.\ Astrophys.\  {\bf 564}, A125 (2014)
  doi:10.1051/0004-6361/201322971
  [arXiv:1402.0004 [astro-ph.HE]].

\bibitem{Fowlie:2016hew} 
  A.~Fowlie and M.~H.~Bardsley,
  Eur.\ Phys.\ J.\ Plus {\bf 131}, no. 11, 391 (2016)
  doi:10.1140/epjp/i2016-16391-0
  [arXiv:1603.00555 [physics.data-an]].

\bibitem{Abdesselam:2016xqt} 
  A.~Abdesselam {\it et al.},
  arXiv:1608.06391 [hep-ex].
  
\bibitem{Akeroyd:2017mhr} 
  A.~G.~Akeroyd and C.~H.~Chen,
  Phys.\ Rev.\ D {\bf 96}, no. 7, 075011 (2017)
  doi:10.1103/PhysRevD.96.075011
  [arXiv:1708.04072 [hep-ph]].



\bibitem{Aaltonen:2011gp} 
  T.~Aaltonen {\it et al.} [CDF Collaboration],
  Phys.\ Rev.\ Lett.\  {\bf 106}, 121801 (2011)
  doi:10.1103/PhysRevLett.106.121801
  [arXiv:1101.4578 [hep-ex]].


\bibitem{Abazov:2010ti} 
  V.~M.~Abazov {\it et al.} [D0 Collaboration],
  Phys.\ Lett.\ B {\bf 695}, 88 (2011)
  doi:10.1016/j.physletb.2010.10.059
  [arXiv:1008.2023 [hep-ex]].

\bibitem{Aaboud:2017buh} 
  M.~Aaboud {\it et al.} [ATLAS Collaboration],
  JHEP {\bf 1710}, 182 (2017)
  doi:10.1007/JHEP10(2017)182
  [arXiv:1707.02424 [hep-ex]].

\bibitem{Khachatryan:2016zqb} 
  V.~Khachatryan {\it et al.} [CMS Collaboration],
  Phys.\ Lett.\ B {\bf 768}, 57 (2017)
  doi:10.1016/j.physletb.2017.02.010
  [arXiv:1609.05391 [hep-ex]].


\bibitem{Alwall:2014hca} 
  J.~Alwall {\it et al.},
  JHEP {\bf 1407}, 079 (2014)
  doi:10.1007/JHEP07(2014)079
  [arXiv:1405.0301 [hep-ph]].

\bibitem{Aaboud:2017zfn} 
  M.~Aaboud {\it et al.} [ATLAS Collaboration],
  JHEP {\bf 1710}, 141 (2017)
  doi:10.1007/JHEP10(2017)141
  [arXiv:1707.03347 [hep-ex]].
  
\bibitem{Dermisek:2014qca} 
  R.~Dermisek, J.~P.~Hall, E.~Lunghi and S.~Shin,
  JHEP {\bf 1412}, 013 (2014)
  doi:10.1007/JHEP12(2014)013
  [arXiv:1408.3123 [hep-ph]].

\bibitem{Aad:2015typ} 
  G.~Aad {\it et al.} [ATLAS Collaboration],
  JHEP {\bf 1603}, 127 (2016)
  doi:10.1007/JHEP03(2016)127
  [arXiv:1512.03704 [hep-ex]].

\bibitem{Aaboud:2016dig} 
  M.~Aaboud {\it et al.} [ATLAS Collaboration],
  Phys.\ Lett.\ B {\bf 759}, 555 (2016)
  doi:10.1016/j.physletb.2016.06.017
  [arXiv:1603.09203 [hep-ex]].

\bibitem{Amhis:2014hma} 
  Y.~Amhis {\it et al.} [Heavy Flavor Averaging Group (HFAG)],
  arXiv:1412.7515 [hep-ex].
  
\bibitem{Buras:2012fs} 
  A.~J.~Buras and J.~Girrbach,
  JHEP {\bf 1203}, 052 (2012)
  doi:10.1007/JHEP03(2012)052
  [arXiv:1201.1302 [hep-ph]].

\bibitem{Alok:2017jgr} 
  A.~K.~Alok, B.~Bhattacharya, D.~Kumar, J.~Kumar, D.~London and S.~U.~Sankar,
  Phys.\ Rev.\ D {\bf 96}, no. 1, 015034 (2017)
  doi:10.1103/PhysRevD.96.015034
  [arXiv:1703.09247 [hep-ph]].

\bibitem{Alok:2017sui} 
  A.~K.~Alok, B.~Bhattacharya, A.~Datta, D.~Kumar, J.~Kumar and D.~London,
  arXiv:1704.07397 [hep-ph].

\bibitem{Mishra:1991bv} 
  S.~R.~Mishra {\it et al.} [CCFR Collaboration],
  Phys.\ Rev.\ Lett.\  {\bf 66}, 3117 (1991).
  doi:10.1103/PhysRevLett.66.3117
  
\bibitem{Borzumati:1998tg} 
  F.~Borzumati and C.~Greub,
  Phys.\ Rev.\ D {\bf 58}, 074004 (1998)
  doi:10.1103/PhysRevD.58.074004
  [hep-ph/9802391].

\bibitem{Sage:2016uxt} 
  F.~S.~Sage, J.~N.~E.~Ho, T.~G.~Steele and R.~Dick,
  arXiv:1611.03367 [hep-ph].

\bibitem{CDF:2012qwd} 
  CDF Collaboration [CDF Collaboration],
  CDF-NOTE-10894.


\bibitem{Aaij:2016flj} 
  R.~Aaij {\it et al.} [LHCb Collaboration],
  JHEP {\bf 1611}, 047 (2016)
  Erratum: [JHEP {\bf 1704}, 142 (2017)]
  doi:10.1007/JHEP11(2016)047, 10.1007/JHEP04(2017)142
  [arXiv:1606.04731 [hep-ex]].
  
  
\bibitem{Chatrchyan:2013cda} 
  S.~Chatrchyan {\it et al.} [CMS Collaboration],
  Phys.\ Lett.\ B {\bf 727}, 77 (2013)
  doi:10.1016/j.physletb.2013.10.017
  [arXiv:1308.3409 [hep-ex]].


\bibitem{Khachatryan:2015isa} 
  V.~Khachatryan {\it et al.} [CMS Collaboration],
  Phys.\ Lett.\ B {\bf 753}, 424 (2016)
  doi:10.1016/j.physletb.2015.12.020
  [arXiv:1507.08126 [hep-ex]].
  



   
  
\bibitem{Lees:2013nxa} 
  J.~P.~Lees {\it et al.} [BaBar Collaboration],
  Phys.\ Rev.\ Lett.\  {\bf 112}, 211802 (2014)
  doi:10.1103/PhysRevLett.112.211802
  [arXiv:1312.5364 [hep-ex]].



\bibitem{Aaij:2014pli} 
  R.~Aaij {\it et al.} [LHCb Collaboration],
  JHEP {\bf 1406}, 133 (2014)
  doi:10.1007/JHEP06(2014)133
  [arXiv:1403.8044 [hep-ex]].
   
  
  
\bibitem{Aaij:2015esa} 
  R.~Aaij {\it et al.} [LHCb Collaboration],
  JHEP {\bf 1509}, 179 (2015)
  doi:10.1007/JHEP09(2015)179
  [arXiv:1506.08777 [hep-ex]].
  
\bibitem{Tolk:2017dqr} 
  S.~Tolk [LHCb Collaboration],
  arXiv:1704.06953 [hep-ex].

\bibitem{Aaij:2015oid} 
  R.~Aaij {\it et al.} [LHCb Collaboration],
  JHEP {\bf 1602}, 104 (2016)
  doi:10.1007/JHEP02(2016)104
  [arXiv:1512.04442 [hep-ex]].
  
\bibitem{CMS:2017ivg} 
  CMS Collaboration [CMS Collaboration],
  CMS-PAS-BPH-15-008.

\bibitem{ATLAS:2017dlm} 
  The ATLAS collaboration [ATLAS Collaboration],
  ATLAS-CONF-2017-023.

\end{thebibliography}
\end{document}